\documentclass{aa}

\usepackage[varg]{txfonts}
\usepackage[colorlinks = true, breaklinks = true, citecolor = blue, linkcolor = blue, urlcolor = blue]{hyperref}
\usepackage[version=4]{mhchem}
\usepackage{bbold}
\usepackage{physics}
\usepackage[separate-uncertainty=true]{siunitx}
\DeclareSIUnit\year{yr}
\DeclareSIUnit\AU{AU}
\DeclareSIUnit\dex{dex}
\usepackage{subcaption}
\usepackage{multirow}
\usepackage{longtable}

\begin{document}

\title{Revisiting equilibrium condensation and rocky planet compositions}

\subtitle{Introducing the \textsc{ECCOplanets} code}

\author{Anina Timmermann\inst{1}
 \and Yutong Shan\inst{3}
 \and Ansgar Reiners\inst{1}
 \and Andreas Pack\inst{2}}

\institute{Institut f{\"u}r Astrophysik und Geophysik, Georg-August-Universit{\"a}t G{\"o}ttingen, Friedrich-Hund-Platz 1, 37077 G{\"o}ttingen, Germany\\
 \email{t.timmermann@stud.uni-goettingen.de}
 \and Geowissenschaftliches Zentrum
Abteilung Geochemie und Isotopengeologie
Goldschmidtstraße 1, 37077 G{\"o}ttingen, Germany\\
 \and Centre for Earth Evolution and Dynamics, University of Oslo, Sem Salands vei 2A ZEB-bygget 0371 Oslo, Norway}

\date{Received XX XX XXXX / Accepted XX XX XXXX}

% \abstract{}{}{}{}{} 
% 5 {} token are mandatory
 
 \abstract
 {The bulk composition of exoplanets cannot yet be directly observed. Equilibrium condensation simulations help us better understand the composition of the planets' building blocks and their relation to the composition of their host star.}
 {We introduce \textsc{ECCOplanets}, an open-source Python code that simulates condensation in the protoplanetary disk. Our aim is to analyse how well a simplistic model can reproduce the main characteristics of rocky planet formation. For this purpose, we revisited condensation temperatures ($T_c$) as a means to study disk chemistry, and explored their sensitivity to variations in pressure ($p$) and elemental abundance pattern. We also examined the bulk compositions of rocky planets around chemically diverse stars.}
 {Our $T$-$p$-dependent chemical equilibrium model is based on a Gibbs free energy minimisation. We derived condensation temperatures for Solar System parameters with a simulation limited to the most common chemical species. We assessed their change ($\Delta T_c$) as a result of $p$-variation between $10^{-6}$ and \SI{0.1}{\bar}. To analyse the influence of the abundance pattern, key element ratios were varied, and the results were validated using solar neighbourhood stars. To derive the bulk compositions of planets, we explored three different planetary feeding-zone (FZ) models and compared their output to an external $n$-body simulation.}
 {Our model reproduces the external results well in all tests. For common planet-building elements, we derive a $T_c$ that is within $\pm\SI{5}{\K}$ of literature values, taking a wider spectrum of components into account. The $T_c$ is sensitive to variations in $p$ and the abundance pattern. For most elements, it rises with $p$ and metallicity. The tested pressure range ($10^{-6}$ -- \SI{0.1}{\bar}) corresponds to $\Delta T_c \approx +\SI{350}{\K}$, and for $-0.3 \leq\left[\ce{M}/\ce{H}\right]\leq0.4$ we find $\Delta T_c \approx +\SI{100}{\K}$. An increase in \ce{C}/\ce{O} from $0.1$ to $0.7$ results in a decrease of $\Delta T_c \approx -\SI{100}{\K}$. Other element ratios are less influential. Dynamic planetary accretion can be emulated well with any FZ model. Their width can be adapted to reproduce gradual changes in planetary composition.}
 {}

\keywords{Planets and satellites: composition -- Planets and satellites: formation -- Protoplanetary disks}

\maketitle 

%
%-------------------------------------------------------------------

\section{Introduction}
The chemical composition of rocky planets is, among other factors such as the surface temperature or the presence of a magnetic field, an important parameter with respect to the habitability and the potential existence of extraterrestrial life. The bulk composition of a planet can be roughly estimated using density measurements from observational data of radial velocity (planet mass) and planetary transits \citep[planet radius;][]{Fulton2018,Zeng2019,Fridlund2020,Otegi2020,Otegi2020a,Schulze2021}. Transit spectroscopy is beginning to provide answers regarding the existence of specific molecules in planetary atmospheres \citep{Seager2000,Brown2001,Madhusudhan2019,Brogi2019,Madhusudhan2021,Rustamkulov2022}. However, the observational answer to the pivotal question of the composition of rocky planets, in particular the makeup of their interior structure, still remains largely elusive. Currently, the only technique that allows glimpses into solids compositions is the spectroscopic analysis of polluted white dwarfs \citep{Jura2014,Farihi2016,Harrison2018,Wilson2019,Bonsor2020,Veras2021,Xu2021}.

This lack of observational data on planetary compositions is currently bridged by simulations of planet formation. To estimate the elemental composition of a planet, it is common to look to its stellar host. The star's chemical composition is assumed to approximately represent that of its protoplanetary disk due to the fact that stars and their disks form from the same molecular cloud \citep{Wang2019,Adibekyan2021}. Out of such a disk, planetesimals and planet embryos form from locally condensed solid materials, whose atomic and mineralogic makeups depend not only on the bulk proportions of elements available, but also on the local temperature and pressure, as governed by the principles of chemical equilibration. In other words, to the first order, the types of building blocks that comprise a planet are determined by the condensation sequence for a given stellar composition and formation location. Due to the assumption of both (1) a chemical equivalence of the molecular cloud and the protoplanetary disk and (2) a chemical equilibrium within the disk, it is immaterial if planetary building blocks truly develop in situ or are inherited from molecular clouds.

The validity of this paradigm has been well tested within our Solar System \citep{Bond2010b,Wang2019}. Information about the composition of the protoplanetary disk of the Solar System has traditionally been inferred from the analysis of meteorites. The carbonaceous chondrites, in particular of the Ivuna-type (CI chondrites), have been found to reflect the elemental abundance of the Sun's photosphere to a very high degree \citep{Lodders2003a,Asplund2009} -- if one allows for some depletion of volatile elements, especially a deficiency in \ce{H}, \ce{C}, \ce{N}, \ce{O}, and noble gases \citep{Righter2006}.

Regarding the rocky planets of our Solar System, it has been found that the composition of the Earth conforms very well to expectations \citep{Wang2019,Schulze2021}: relative to the composition of the Sun (measured by spectroscopy and approximated by the CI chondrites), the bulk Earth is depleted in moderately volatile elements, that is, elements that condense into mineral dust grains at lower temperatures \citep{Kargel1993,Carlson2014,Wood2019,Wang2019}, but reflects the Sun's elemental abundance pattern for refractory elements. Much less is known about the composition of the other rocky planets of the Solar System. There are several processes that can modify the bulk composition of a planet. For instance, the planetary material may be taken out of the chemical equilibrium of the disk at some threshold temperature (`incomplete condensation'), as was likely the case for Earth \citep{Wood2019,Sossi2022}. Another example is Mercury, with its high density. Mercury has lost about 80 percent of its rocky (low density) mantle via collisional erosion \citep{Benz1988}. In contrast to Mercury, the bulk density of Mars is lower than expected when assuming a CI chondritic bulk composition \citep{Schulze2021}. These findings show the limits of predicting planetary bulk compositions based on the element abundance of their host star alone. Complex processes, such as dynamical interactions, migration, radial mixing, discontinuous distribution of solids, or even giant impact events can offset a planet's abundance pattern \citep[see e.g.][]{Benz1988,Clement2021,Izidoro2022}. The identification of such anomalies, however, requires knowledge of the expected bulk density, which can only be obtained from the bulk chemistry and size of a planet.

It is now becoming possible to investigate how well exoplanet systems conform to this picture, at least indirectly. \citet{Schulze2021} explored the statistical likelihood of rocky planets having the same composition as their host stars, by comparing their expected core mass fraction to the core mass fraction that could be expected given the elemental abundance of the star. Of their sample of eleven planets, only two were found to be incompatible with the null-hypothesis assumption of the planet reflecting the composition of its host star at the $1\sigma$ level. Similarly, \citet{Plotnykov2020} analysed an ensemble of planets and stars and find that the predicted composition of the population of planets spans an overlapping, yet wider, range with respect to the corresponding host stars, both in terms of the \ce{Fe}/\ce{Si} distribution and the core mass fraction \citep[see also][]{Adibekyan2021}.

Given these findings, we can take advantage of the possibility to deduce a star's composition from its spectrum and assume the same elemental ratios for the protoplanetary disk. The advances in modern spectrographs, coupled with an improved understanding of spectral line characteristics, allows the derivation of precise abundances of major rock-building elements, at least for F, G, and K stars \citep{Adibekyan2012,Brewer2016}. Despite all the advances in this field, it should be kept in mind that deducing concrete stellar elemental abundances from the depth and width of absorption features in a spectrogram is far from straightforward. As compiled and analysed by \citet{Hinkel2016}, there are a multitude of methods that obtain quite different elemental abundances with different error margins, even from the same spectra \citep[see also][]{Bedell2014}. 

Using simulations to find the composition of exoplanets depending on the composition of their host star has become a fairly common practice. Apart from trying to recreate the planets of the Solar System in order to test our understanding of planet formation \citep{Raymond2004,OBrien2006,Bond2010b}, there have been great efforts to explore the compositional diversity of exoplanets \citep{Bond2010a,Johnson2012,Thiabaud2015,Dorn2019}, and especially the influence of stellar elemental abundance patterns that deviate significantly from that of our Sun \citep{Bitsch2020,Carter-Bond2012,Jorge2022}.

Due to its provision of an extensive thermochemical database and its widely applicable general chemical equilibrium computations, the powerful commercial software suite \textsc{HSC Chemistry}\footnote{\href{http://www.hsc-chemistry.net}{http://www.hsc-chemistry.net}} has been the backbone of many recent studies of exoplanet compositions \citep[for instance,][]{Bond2010b,Johnson2012,Thiabaud2014,Moriarty2014,Thiabaud2015,Dorn2019}. There are also several general equilibrium condensation codes that were written specifically for applications in planetary science but which are generally not publicly available, such as the \textsc{Condor} code, developed by and described in \citet{Lodders1993}, and the \textsc{PHEQ} code, developed by and described in \citet{Wood1993}. A freely available Fortran code is \textsc{GGchem}  \citep{Woitke2018}. This code simulates the equilibrium chemistry in the protoplanetary disk down to \SI{100}{\K} and includes an extensive thermochemical database. Another open-source program is the \textsc{TEA} code by \citet{Blecic2016b}, which, however, is limited to gas-phase simulations. Originally intended to study geochemical processes, but also usable for planetary simulation, is the \textsc{SUPCRTBL} software package by \citet{Zimmer2016}, with its extensive thermochemical database \textsc{SUPCRT92}. 

Building on the foundation of results from general equilibrium calculations, additional effects can be included to capture the complexity of the disk evolution process. Examples include: combining a thermochemical equilibrium simulation with a dynamical simulation of disk development \citep{Bond2010b,Bond2010a,Moriarty2014,Thiabaud2015,Khorshid2022}; including dust enrichment in the composition of the protoplanetary disk to account for deviations of planet compositions from stellar elemental ratios \citep{Ebel2000a}; adding the notion of the isolation of a fraction of the condensed solids from the chemical equilibrium \citep{Petaev1998a}; and looking into non-ideal solid solutions \citep[e.g.][]{Pignatale2011}. Once the most likely composition of a rocky planet has been found, further simulations can follow to estimate the internal structure of the planet \citep[see e.g.][]{Rodriguez-Mozos2022}, its geological evolution \citep{Putirka2021}, the development of an atmosphere 
\citep{Herbort2020,Spaargaren2020,Ballmer2021,Putirka2021}, and even the formation and composition of clouds \citep{Herbort2022}.

With our \textsc{ECCOplanets}\footnote{\textbf{E}quilibrium \textbf{C}ondensation \& \textbf{C}omposition \textbf{O}f \textbf{planets}, or in Italian roughly `There you have it: planets'. Available at \href{https://github.com/AninaTimmermann/ECCOplanets}{https://github.com/AninaTimmermann/ECCOplanets}.} code, we provide a general equilibrium condensation code as a simplified, open-source Python alternative to use for simulations. The main focus of our code is its ease of use, which allows it to be tailored to specific research questions and extended. As a first application of our code, we show its projection for the composition of exoplanets in different stellar systems and the condensation temperatures of common planet-building molecules and elements. We also study the sensitivity of condensation temperatures to variations in disk pressure and elemental abundance patterns within the protoplanetary disk. With this analysis, we want to highlight the limitations of the application of element volatility, as determined from Solar System parameters, in general theories of planet formation.

In Sect. \ref{sec:ThermChemBasis} we describe the underlying thermochemical principles of our simulation. Section \ref{sec:MathAnal} deals with the mathematical properties of the thermochemical equations. Our data sources and processing are presented in Sect. \ref{sec:Data}. The basics of our code are shown in Sect. \ref{sec:CompSol}. Finally, we present our simulation results regarding condensation temperatures of certain species and their variability (Sect. \ref{sec:SimRes}) and the composition of exoplanets (Sect. \ref{sec:SimRes_PComp}). Apart from showing our own results, these simulations are used as a benchmark test for our code.

%%%%%%%%%%%%%%%%%%%%%%%%%%%%%%%%%%%%%%%%%%%%%%%%%%%%%%%%%%%%%%%%%%%%%%%%%%%%%%%%%%%%%%%%%%%%%%%%%%%%%%%%%%%%%%%%%%%%%%%%%%%%%%%%%%%%%%%%%%%%%%%%%%%%%%%%%%%%%%%%%%%%%%%%%%%%%%%%%%%%%%%%%%%%%%%%%%%%%%
\section{Thermochemical basis}\label{sec:ThermChemBasis}
A protoplanetary disk, at least during the stage of solid condensation, can be approximated as a closed system, that is, closed with respect to matter but open with respect to heat exchange. The timescale on which the disk cools is generally large compared to the timescale of condensation \citep{Toppani2006,Pudritz2018}. This is, however, only true for the formation of condensates from the gas phase, not for the rearrangement of the condensates into their thermochemically favoured phases \citep[for estimates of the relevant timescales, see][]{Herbort2020}. Thus, assuming that the disk evolves through a sequence of equilibrium conditions is, to some degree, a simplification, especially for large bodies at low temperatures. 

\subsection{Chemical equilibrium and Gibbs free energy minimisation}
A system is in thermochemical equilibrium when its Gibbs free energy is minimised \citep{White1958,Eriksson1971}. Accordingly, we can compute the disk's equilibrium composition at each temperature by minimising its Gibbs free energy.

The Gibbs free energy is an extensive property, that is, the total Gibbs energy of a multi-component system is given by the sum of the Gibbs energies of its constituents \citep{Eriksson1971}:
\begin{equation}\label{eq:G_tot}
 G_{\rm{tot}}(T) = \sum_i G_i(T).
\end{equation}

The Gibbs energy of a substance $i$ is given by its molar amount $x_i$ and its chemical potential $\mu_i(T)$, also referred to as its molar Gibbs free energy
\citep{Eriksson1971}:
\begin{equation}\label{eq:G_i}
G_i(T) = x_i\; \mu_i(T).
\end{equation}

The chemical potential at a temperature $T$ depends on the chemical potential at standard state, $\mu_i^\circ(T)$, and the natural logarithm of the chemical activity, $a_i$, of the component \citep{Eriksson1971}
\begin{equation}\label{eq:g_i}
\mu_i(T) = \mu_i^\circ(T) + R\,T\ln{a_i},
\end{equation}
where $R$ is the ideal gas constant.

Regarding the first term of Eq. (\ref{eq:g_i}), we use the relation of the standard Gibbs free energy, $G^\circ$, to the standard enthalpy, $H^\circ$, and the standard entropy, $S^\circ$:
\begin{equation}
 \dd G^\circ = \dd H^\circ - T\,\dd S^\circ.
\end{equation}
If we use these variables as molar quantities, that is, enthalpy and entropy per mole of a substance, this equation holds for the standard chemical potential, $\mu^\circ(T)$ \citep[cf.][Ch. 8.3]{Keszei2012}.

The dependence of the enthalpy and entropy on changes in temperature is defined by the heat capacity, $C_p^\circ$ \citep[cf.][Chs. 4.4.1, 4.4.2]{Keszei2012}:
\begin{align}
    \dd H^\circ &= C_p^\circ(T)\; \dd T\\
    \dd S^\circ &= \frac{C_p^\circ(T)}{T}\; \dd T.
\end{align}

The heat capacity, $C_p^\circ$, is well approximated by the Shomate polynomial \citep{Chase1998,NISTwebbook}, allowing us to integrate the differentials analytically. With $\tau = T \times10^{-3}$ it takes the form
\begin{equation}
 C_p^\circ(T) = A + B\,\tau + C\,\tau^2 + D\,\tau^3 + E\,\tau^{-2}.
\end{equation}

The Shomate equation is only valid for temperatures larger than $T=\SI{298.15}{\K}$. This is also the reference temperature of the standard state of all thermochemical data used in our code. Therefore, it constitutes the lower limit of the integration. The upper limit is given by any temperature, $T$: 
\begin{align}
 H^\circ(T) & - H^\circ(\SI{298.15}{\K}) = \int_{\SI{298.15}{\K}}^T C_p^\circ(T) \dd T\\
 &= A\,\tau + \frac{1}{2}\,B\,\tau^2 + \frac{1}{3}\,C\,\tau^3 + \frac{1}{4}\,D\,\tau^4 - E\,\tau^{-1} + F \label{eq:Shomate_H0}\\
 S^\circ(T) &- S^\circ(\SI{298.15}{\K}) = \int_{\SI{298.15}{\K}}^T \frac{C_p^\circ(T)}{T} \dd T\\ 
 &= A\,\ln{(\tau)} + B\,\tau + C\,\frac{\tau^2}{2} + D\,\frac{\tau^3}{3} - \frac{E}{2\,\tau^2} + G \label{eq:Shomate_S0},
\end{align}
where $F$ and $G$ denote the negative value of the integrated polynomials evaluated at $T=\SI{298.15}{\K}$. We can rearrange the equations and add the constants $H^\circ(\SI{298.15}{\K})$ and $S^\circ(\SI{298.15}{\K})$ to the constants $F$ and $G$, respectively, on the right-hand side. The new constants are denoted with a tilde sign.\footnote{This is a slight deviation from the definition of the constants in the NIST-JANAF web-book \citep{NISTwebbook}.} 

Combining the resulting polynomials for $H^\circ(T)$ and $S^\circ(T)$ gives us an equation for the standard chemical potential of each species defined by the Shomate parameters:\begin{small}
\begin{align}
 \mu_i^\circ(T) & =H_{i}^\circ(T) - T\,S_{i}^\circ(T)\\
 &= 10^3\,\left[A\,\tau\,(1-\ln{\tau}) - \frac{B\,\tau^2}{2} - \frac{C\,\tau^3}{6} - \frac{D\,\tau^4}{12} - \frac{E}{2\,\tau} + \widetilde{F} - \widetilde{G}\,\tau \right].
\end{align}
\end{small}

The usage of $\tau = T \times10^{-3}$ entails that for consistent constants $A$ to $G$, the enthalpy, $H^\circ$, is given in units of [\si{\kJ\per\mol}], whereas the heat capacity, $C_p^\circ$, and the entropy, $S^\circ$, are given in [\si{\J\per\mol\per\K}]. The chemical potential, $\mu^\circ(T)$, is given in [\si{\J\per\mol}].

In this version of the code, we treat the gas phase as ideal gas and only consider pure solid phases, that is, no solid solutions. We can therefore use a common approximation for the activity of a substance \citep[second term of Eq. (\ref{eq:g_i}); see e.g.][]{Eriksson1971}: For solids, the activity is unity; for components in the gas phase, the activity is assumed to equal the partial pressure of this component. This also entails that we do not differentiate between stable and unstable condensed phases on the basis of the activity. The presence of a phase is solely determined by the product of its molar amount and chemical potential at standard state. The gas-phase approximation is valid, as long as deviations from an ideal gas are negligible. This deviation can be quantified with the fugacity coefficient, which depends on the gas in question, the pressure, and the temperature. As a rule of thumb, this approximation of an ideal gas is better the lower the pressure and the higher the temperature \citep[see e.g.][Ch. 1.2]{Atkins2006}, we do not expect significant influences on our result for our parameter range where $T > \SI{300}{\K}$ and $p < \SI{1}{\bar}$.

The partial pressure of a component, $p_i$, can be expressed as the product of the total pressure with the fraction of the molar amount of the species in question, $x_i$, of the total molar amount of all gaseous species, $X$. In our case the total pressure is that of the protoplanetary disk $p_{\rm{disk}} = p_{\rm{tot}} := p$. The activity can be summed up as 
\begin{equation}\label{eq:a_i}
 a_{i} = 
 \begin{cases}
 p_i = \frac{x_i}{X}\; p  & \text{gas-phase species},\\
 1 & \text{solid-phase species}.
 \end{cases}
\end{equation}

By combining Eqs. (\ref{eq:G_tot}) to (\ref{eq:a_i}), and assembling all molar amounts $x_i$ in the vector $\mathbf{x}$, we get the Gibbs free energy function of the protoplanetary disk that needs to be minimised in order to find the equilibrium composition \citep{Eriksson1971}:
\begin{align}\label{eq:G_tot_therm}
 G_{\rm{sys}}(\mathbf{x},T,p) &= \sum_i x_i \left[\mu_i^\circ + R\,T\,\ln{a_i}\right]\\
 &= \sum\limits_{\rm{gas},i} x_i\left[\mu_i^\circ + R\,T\,\right(\ln{p} + \ln{\frac{x_i}{X}} \left)\right] + \sum\limits_{\rm{solid},i} x_i\,\mu_i^\circ.
\end{align}

This equation is a function of the molar amounts $x_i$ of all chemical species in the system. The temperature $T$ and disk pressure $p$ can be specified or sequentially varied, all other factors are constants. Accordingly, minimising Eq. (\ref{eq:G_tot_therm}) means finding the molar amounts $x_i$ of all chemical species in chemical equilibrium.

\subsection{Constraints}
There are two constraints in the minimisation of the Gibbs free energy. The most obvious constraint to our problem is non-negativity, which means that no molar amount can take values below zero:
 \begin{equation}
    x_i \geq 0\; \forall \;x_i \in \text{System}. 
 \end{equation}

The second constraint, mass balance, is imposed by the elemental composition of the protoplanetary disk. The amount of any element bound in molecular species cannot exceed the initial molar amount of that element. If we include all possible forms in which an element can occur, we can formulate this requirement as an equality constraint:
  \begin{equation}
      \sum\limits_i \alpha_{ij}\,x_i = \beta_j,
 \end{equation}
where $\beta_j$ is the initial molar amount of element $j$ and $\alpha_{ij}$ is the stoichiometric number of element $j$ in species $i$ (see the example in Appendix \ref{app:stoichiometry}). It should be noted that our code only considers relative amounts of species, based on relative initial abundance patterns. We specify all molar amounts relative to an initial abundance of $10^6$ \ce{Si} atoms. 

%%%%%%%%%%%%%%%%%%%%%%%%%%%%%%%%%%%%%%%%%%%%%%%%%%%%%%%%%%%%%%%%%%%%%%%%%%%%%%%%%%%%%%%%%%%%%%%%%%%%%%%%%%%%%%%%%%%%%%%%%%%%%%%%%%%%%%%%%%%%%%%%%%%%%%%%%%%%%%%%%%%%%%%%%%%%%%%%%%%%%%%%%%%%%%%%%%%%%%
\section{Mathematical analysis}\label{sec:MathAnal}

We characterise the problem as a constrained, non-linear optimisation of the general form \citep[see e.g.][Ch. 4.1]{Boyd2004}:
\begin{align}
 \text{minimise} &\qquad f(x)\\
 \text{subject to} &\qquad g(x) \geq 0\\
 &\qquad h(x) = 0,
\end{align}
where the target function $f(x)$ is the function of the Gibbs free energy of the system, $g(x)$ is the non-negativity constraint and $h(x)$ is the mass balance constraint.

The target function $f(x)$ is mathematically the most complex of the three functions. As shown in the previous section, for $n$ gas-phase species and $m$ solid-phase species, it can be written as
\begin{equation}\label{eq:G_tot_math}
 f:\mathbb{R}^{n+m} \rightarrow \mathbb{R}, \textbf{x} \rightarrow G_{\rm{sys}}(\mathbf{x},T,P) = 
\underbrace{\textbf{c}^T\,\textbf{x}}_{\text{linear}} + \underbrace{d\,\sum\limits_{i=1}^{n} x_i \, \ln{\frac{x_i}{X}}}_{\text{transcendental}},
\end{equation} 
with $\textbf{c},\;\textbf{x} \in \mathbb{R}^{n+m}$ and $d \in \mathbb{R}$,

\begin{equation}\label{eq:const_c_d}
 \textbf{c} = \begin{pmatrix}\mu_1^\circ + R\,T\,\ln{p}\\\vdots\\\mu_n^\circ + R\,T\,\ln{p}\\ \mu_{n+1}^\circ\\\vdots\\\mu_{n+m}^\circ\end{pmatrix} \hspace{2cm}
 d = R\,T.
\end{equation} 

As denoted, the function is the sum of a linear and a transcendental function. For the further characterisation of the transcendental term, we use the fact that it is the sum of terms that are identical in mathematical form, 
\begin{equation}
 f_{\rm{trans, i}}(x_i) = d\, x_i \, \ln{\frac{x_i}{X}}.
\end{equation} 
The molar amount $x_i$ of any species $i$ has to be between 0 and the total molar amount of all species $X$. For computational purposes, we use the limit and define
\begin{equation}
 \lim\limits_{x \to 0} x\,\ln{x} = 0 \overset{\text{define}}{\longrightarrow} 0\,\ln{0} := 0.
\end{equation}
With this definition, $f_{\rm{trans, i}}(x_i), x_i \in [0,X]$, is U-shaped and has zero intercepts at zero and $X$ only. It is convex over its domain. The total transcendental term is the sum of convex functions, and thus convex itself \citep[cf.][Ch. 3.2]{Boyd2004}. Adding the linear term has no influence on this characterisation. So, the target function is convex. The main implication of the convexity of a function is that finding a local minimum is always equivalent to finding a global minimum \citep[cf.][Ch. 4.2.2]{Boyd2004}, that is, our minimisation solution is unique.

The non-negativity constraint can be phrased as an inequality constraint:
\begin{align}\label{eq:ineq}
 \mathbf{x} \geq \mathbf{0} \hspace{1cm} \text{or} \hspace{1cm} g(\mathbf{x}) = \mathbf{x} \geq 0,
\end{align}
with $\mathbf{x}$ as defined above.

The equality constraint can be written as
\begin{align}
 \mathbf{A}\,\mathbf{x} = \mathbf{b} \hspace{1cm} \text{or} \hspace{1cm} h(\mathbf{x}) = \mathbf{A}\,\mathbf{x} - \mathbf{b} = 0,
\end{align}
where $\mathbf{A} \in \mathbb{N}^{o\times p}$ is the stoichiometry-matrix for the number balance of a system composed of initially $o$ elements resulting in a final composition with $p = m+n$ molecular species (see the example in Appendix \ref{app:stoichiometry}), and $\mathbf{b} \in \mathbb{R}^{o}$ is the vector containing the total abundances of the elemental components. Typically, $p > o$ because the most common molecular species are only made up of a handful of different elements.

The gradient of the target function is given by
\begin{align}\label{eq:dG_dx}
 \frac{\partial f}{\partial x_i} = 
 \begin{cases}
 c_i + d\,\ln{\frac{x_i}{X}} & i \leq n,\; x_i \neq 0\\
 c_i & i > n \: \text{or}\: x_i = 0.\\
 \end{cases}
\end{align}
\noindent
The Hesse matrix of the target function is given by 
\begin{align}\label{eq:d2G_dxdy}
 \frac{\partial^2 f}{\partial x_i \partial x_j} = 
 \begin{cases}
 -\frac{d}{X} & i,j \leq n,\: i \neq j \: \text{or}\: x_i = 0,\\
 \frac{d}{x_i} - \frac{d}{X} & i,j \leq n,\: i = j,\; x_i \neq 0,\\
 0 & \text{else}.
 \end{cases}
\end{align}

In summary, our problem can be characterised as a convex minimisation problem subject to two linear constraints. For the numerical minimisation we can make use of the gradient and Hesse matrix of the target function, and can be sure of the uniqueness of a found solution due to the convexity of the target function.

%%%%%%%%%%%%%%%%%%%%%%%%%%%%%%%%%%%%%%%%%%%%%%%%%%%%%%%%%%%%%%%%%%%%%%%%%%%%%%%%%%%%%%%%%%%%%%%%%%%%%%%%%%%%%%%%%%%%%%%%%%%%%%%%%%%%%%%%%%%%%%%%%%%%%%%%%%%%%%%%%%%%%%%%%%%%%%%%%%%%%%%%%%%%%%%%%%%%%%
\section{Data}\label{sec:Data}
\subsection{Data types and sources}\label{sec:datatypes_sources}
There are different types of data used in this code. Regarding their function within the code, we can distinguish the thermochemical data of molecules (this term includes minerals; melts were not considered), on the one hand, and the stellar elemental abundance data, on the other hand. In terms of their mathematical usage, the former plays the main role in the target function of the minimisation procedure, whereas the latter is used in the number balance constraint. The thermochemical data describes laboratory-measured properties of molecules and is constant for all simulations; the stellar abundance data were derived from astronomical observations of particular stars and can be varied as an input parameter between simulations. We use ancillary atomic weight data to express the atomic composition of planets in terms of $\text{wt}-\%$.

Our thermochemical database is limited to the most common species expected to form in a protoplanetary disk, and does not contain any charged species at the moment. It is likely that the lack of certain molecules and ions increases the expected error of the computed condensation temperatures, especially for high-$T$ condensates containing \ce{Mg} and \ce{Na}. For the sake of formal comparability between the thermochemical data of different species (i.e. identical derivation, processing, and presentation of data), we used as few different data sources as possible. Most data, especially the gas-phase data, were taken from the comprehensive NIST-JANAF Thermochemical Tables.\footnote{\href{https://janaf.nist.gov/}{https://janaf.nist.gov/}} Most of the mineral data were taken from three bulletins of the U.S. Geological Survey \citep{Robie1979,Robie1995,Hemingway1982}. The data were extracted from these sources in their given tabulated form. An overview of the included data is shown in Appendix \ref{app:molecules}.

The stellar elemental abundance data are given as the absolute number of atoms of each element, normalised to $N_{\ce{Si}} = 10^6$. This is an arbitrary scaling commonly used in cosmochemistry \citep[see e.g.][]{Lodders2003a,Bond2010a,Lodders2020}. The exact normalisation is inessential for the code, as we only consider the element ratios in the disk. We included the data of $1617$ F, G, and K stars from the 
\citet{Brewer2016} database in the code.

Further data can easily be added to all databases if considered useful for a simulation.

\subsection{Data processing, uncertainties, and extrapolation}\label{sec:dataprocess}
The tabulated data are only available at discrete temperatures, with intervals of \SI{100}{\K}. To obtain continuous thermochemical data for any temperature, we used the tabulated enthalpies to fit Shomate parameters $A$, $B$, $C$, $D$, $E$, and $F$, via the respective Shomate equation (cf. Eq. (\ref{eq:Shomate_H0})). Subsequently, we use the found values ($A$ to $E$), the tabulated entropies and the respective Shomate equation (Eq. (\ref{eq:Shomate_S0})) to find the last parameter $G$.

\begin{figure}[ht]
 \centering
 \includegraphics[width=0.5\textwidth]{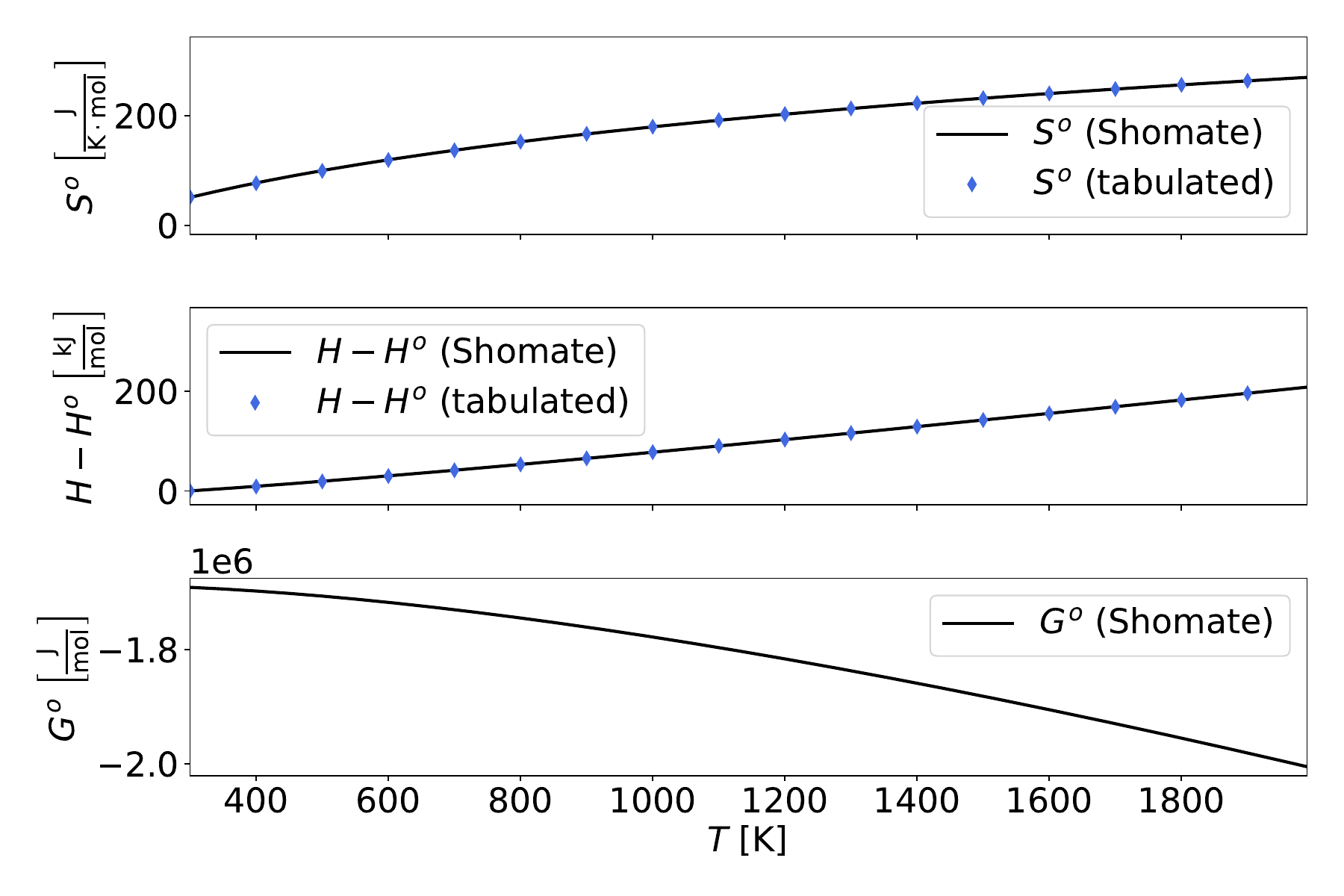}
 \caption{Example of graphical output of the molecule database for corundum (\ce{Al2O3(s)}). \textit{Top panel}: Entropy. \textit{Middle panel}: Enthalpy. \textit{Bottom panel}: Gibbs energy (chemical potential). Blue diamonds show tabulated values (upper two panels) and solid black lines the corresponding fitted Shomate function.}
 \label{fig:mol_db_Al2O3}
\end{figure}

As an example, we show the thermochemical data of \ce{Al2O3(S)} in Fig. \ref{fig:mol_db_Al2O3}. The top and middle panel show the entropy and enthalpy values of the species, respectively. We see that our Shomate fit (black solid line) retraces the tabulated data (blue diamonds) well. The bottom panel shows the Gibbs free energy derived from the other two properties. 

In cases where the tabulated data contains discontinuities in the form of jumps at certain temperatures (commonly in reference-state data including multiple phases of a species), we fitted separate Shomate parameters for either side of the discontinuity. If the tabulated data do not span our entire temperature range of \si{300} to \SI{6000}{\K}, we applied a linear extrapolation for the Shomate enthalpy equation (parameters $A$ and $F$) using the last three tabulated enthalpy values. The $G$ parameter was fitted subsequently.

The uncertainty in thermochemical data can be very large, especially for solids. The reason for this is a combination of the limited precision of experimental measurements and errors introduced in the subsequent data processing, for instance by extrapolating measured data beyond the temperature range of the experiment. Regarding the measurement uncertainty, our data sources state an uncertainty of the order of 10\% for the entropy at reference state for many minerals; for gases the uncertainty is generally much lower, often below 1\% (for the exact values, we refer to the data sources themselves, Sect. \ref{sec:datatypes_sources}). This uncertainty is propagated to the other values in the tables. Regarding the data processing uncertainty, \citet{Worters2018} and \citet{Woitke2018} studied the deviations between the thermochemical equilibrium constants of the species ($k_p$) as stated in different data collections as a function of temperature. They found that only for ~65\% of the species the agreement between different sources is good over the entire temperature range, that is, better than \SI{0.1}{dex} at high temperatures and better than \SI{0.4}{dex} at low temperatures.

We did not take the uncertainty in the thermochemical data into account in the assembling of our thermochemical database, and it is not considered in any way in our simulations. It should, therefore, be kept in mind for the interpretation of our simulation results.

There are also uncertainties in the stellar elemental abundance data. For most rock-building elements, the estimated error of the given abundances is smaller than $\pm\SI{0.03}{dex}$ for the stars in our database \citep{Brewer2016}. We discuss the implications of these uncertainties in Sect. \ref{sec:condT_var_imp}.

%%%%%%%%%%%%%%%%%%%%%%%%%%%%%%%%%%%%%%%%%%%%%%%%%%%%%%%%%%%%%%%%%%%%%%%%%%%%%%%%%%%%%%%%%%%%%%%%%%%%%%%%%%%%%%%%%%%%%%%%%%%%%%%%%%%%%%%%%%%%%%%%%%%%%%%%%%%%%%%%%%%%%%%%%%%%%%%%%%%%%%%%%%%%%%%%%%%%%%
\section{Computational solution}\label{sec:CompSol}
Our computational solution to the chemical equilibrium problem is provided as an open-source software on \textsc{GitHub}\footnote{\href{https://github.com/AninaTimmermann/ECCOplanets}{https://github.com/AninaTimmermann/ECCOplanets}}. It is written in Python and only relies on the standard Python libraries \texttt{numpy}, \texttt{scipy} and \texttt{pandas}.\footnote{We acknowledge the fact that Python is not a particularly efficient coding language (see our performance test in Table \ref{tab:runtime} and compare to e.g. \citet{Woitke2018}, who use Fortran-90 in their \textsc{GGchem}-code), and understand the growing concern about the ecological impact of computational astrophysics \citep{PortegiesZwart2020}. Our choice is motivated by the conjecture that Python is most commonly taught and used in physics, astronomy, and geosciences, and our aim to also make our code accessible to an audience with limited coding experience.} Both the databases and minimisation code can be easily expanded to include more species or integrate a more sophisticated scientific approach, for instance by including an isolation fraction for the simulated solids or a thermochemical activity model.

\subsection{Scope of our simulation}
We limited the scope of our simulation in several areas. Most importantly, it is a purely thermochemical simulation. We do not consider disk profiles, disk dynamics, dynamical planet formation models or planet migration. The temporal development of the disk is only considered indirectly, in that we simulate a decrease in temperature, but do not set an absolute timescale for its evolution. The thermochemical approach is limited to equilibrium condensation, that is, we assume that all components of the system stay in chemical equilibrium for the whole temperature range of $6000$ to $300$ \si{\K}, or, in any case, down to the temperature at which the planet is extracted from the disk. This is a twofold approximation: firstly, we assume that the cooling timescale of the disk is large compared to the thermochemical equilibration timescale, and secondly, we assume that no part of the condensates becomes isolated from the equilibration process, for instance, by being integrated into larger bodies. Furthermore, our model only includes ideal gases and solids. We do not include the condensation of trace elements. 

Our scientific objective is the analysis of the composition of rocky planets; thus, it is limited to the solid materials found after condensation. While we do simulate the evolution of gas-phase species in the disk, they are not considered to be part of a forming planet. We only look at relative amounts of species and make no assumptions regarding absolute planet sizes.

\subsection{Condensation simulation}\label{sec:func_condsim}
The code's main function is to compute the temperature-dependent equilibrium composition of a protoplanetary disk and to allow for the subsequent analysis of the results. This is realised by performing a sequence of Gibbs Free Energy minimisations at decreasing temperatures. As a result, the minimisation routine returns the relative amounts of each species -- gaseous and solid -- at each temperature step. This result matrix allows the inference of the temperature ranges in which a species is stable, and, therefore, expected to be present in the protoplanetary disk.

The code requires the specification of a start and end temperature, as well as the temperature increment, the disk pressure, the elemental abundance pattern in the disk, and the list of molecular species to be considered. The start and end temperature, as well as the temperature increment have no physical meaning; their computational impact is discussed in Appendix \ref{app:StabPerf}. 

The disk pressure is kept constant for the entire simulation, which means it is not automatically adapted to a decrease of gaseous material or temperature. Its value should be chosen in accordance with disk profiles. The elemental abundance pattern needs to be specified in terms of the normalised absolute number of atoms per element (see Sect. \ref{sec:datatypes_sources}).

\subsubsection{Molecule selection}\label{sec:StabPerf_molsel}
A meaningful selection of the species to be included in the simulation is the most important, albeit most complex, aspect of the simulation. Our ideal is to include as few species as possible, in order to make the minimisation problem as numerically easy to solve as possible, thereby reducing the expected computation time and the likelihood of numerical errors. On the other hand, one wishes to find a stable mineral assemblage and hence would like to include all phases where thermodynamic data are available. Currently, there are $>5000$ minerals known to occur in nature; including all would likely lead to computational problems. The difficulty is determining the set of crucial species, that is, those that have a notable influence on the thermochemistry of the disk, for instance, by controlling the total pressure or otherwise determining condensation temperatures.

The total pressure is generally controlled by \ce{He} and \ce{H2}. On top of that, the simulation results will only be reliable if we include the most stable species at any sampled ($T$,$p$)-tuple for each element contained in the simulation. The most stable species carrying a particular element at a given ($T$,$p$)-value depends on the overall elemental abundance pattern; thus, compiling the list of species is typically an iterative process. The starting point of this process can be based on the extensive simulations of the Solar System, for instance by \citet{Lodders2003a}, \citet{Woitke2018}, and \citet{Wood2019}.

\subsubsection{Minimisation procedure}
The minimisation at each temperature step is done using a trust region method, due to its known stability when solving bounded and constrained non-linear problems
\citep{Conn2000}. The Gibbs function $G(\textbf{x},T, p)$, as defined in Eq. (\ref{eq:G_tot_therm}), is passed to the minimisation function as the target function, in combination with the non-negativity and number-balance constraints, the gradient (Eq. (\ref{eq:dG_dx})) and Hessian matrix (Eq. (\ref{eq:d2G_dxdy})) in their respective capacity. 

We derive the initial starting point $\textbf{x}_0$ for the minimisation at the highest temperature of the simulation by solving the linear part of the Gibbs equation (Eqs. (\ref{eq:G_tot_math}), (\ref{eq:const_c_d})) with a simplex method. We ignore the transcendental part of the function, but respect non-negativity and number-balance. 

The vector $\textbf{c}$, as defined in Eq. (\ref{eq:const_c_d}), is constructed from the Shomate parameters of all included species, the specified temperature, and disk pressure. In all subsequent temperature steps, the solution at the previous temperature step is used as the initial guess input of the minimisation procedure.

\subsection{Simulation analysis and definition of the condensation temperature} \label{sec:CompSol_Funcs_SimAnal}

\begin{figure*}[ht]
 \centering
 \includegraphics[width=0.9\textwidth]{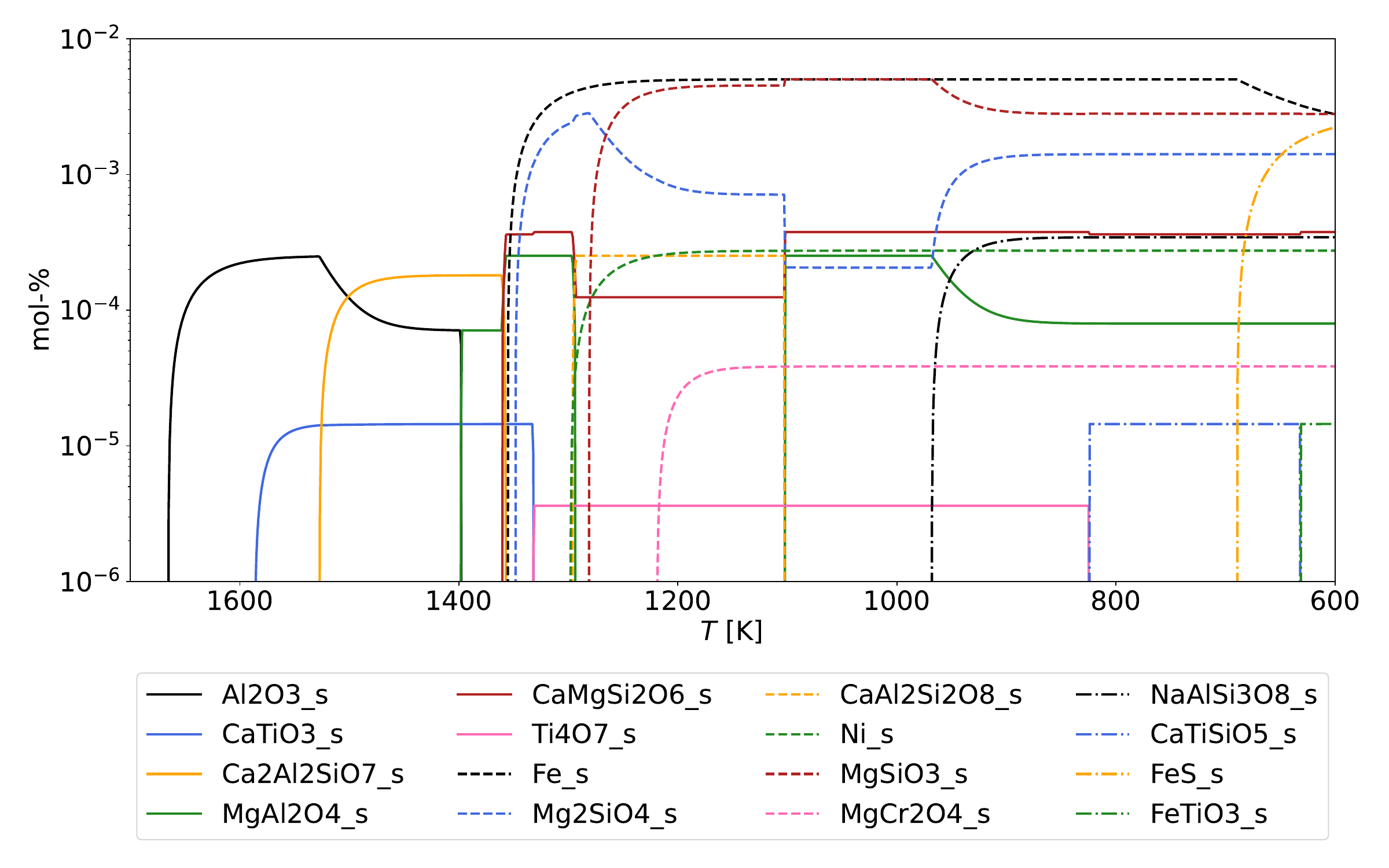}
 \caption{Example of the progression of solid species of a simulation. The molar amount relative to the total molar content of the disk of each of the condensates included in the simulation is shown as a function of decreasing temperature in different line colours and styles, as denoted in the legend. See Appendix \ref{app:molecules} for details on the included species. The simulation was run at a constant disk pressure of $p=10^{-4}~\si{bar}$ and is based the solar elemental abundance pattern recommended by \citet{Lodders2003a}. See Table \ref{tab:SimParams} for a summary of the simulation parameters.}
 \label{fig:plot_s}
\end{figure*}

We provide several functions to analyse the results of the condensation simulation. These analysis functions roughly fall into three categories. First, the basic parameters of the simulation (temperature range, included molecules, abundance pattern, and disk pressure) can be retrieved. Second, temperature progression curves of species can be plotted to analyse their amounts as a function of temperature and, in the case of solid species, their condensation behaviour. And third, the projected composition of a rocky planet as a function of its formation temperature can be studied. 

Temperature progression curves are a powerful tool for analysing different aspects of planet formation. As an example, we show a plot of the relative molar amounts ($\si{mol}-\%$)\footnote{The specification `\si{mol}' is used to distinguish it from the later used weight percentage ($\text{wt}-\%$).} of all the solid species included in a simulation as a function of decreasing temperature in Fig. \ref{fig:plot_s}. The amounts are relative to the total molar content of the disk, that is, both gas- and solid-phase species $\left(\dfrac{n_{\rm{species}}}{n_{solids} + n_{gases}} \text{ in } \si{\mol}-\%\right)$. From this figure, we can gather various information. For instance, the temperature of the first onset of condensation, the approximate total proportion of solids in the disk as a function of temperature, and the dominant contributors to a planet's bulk composition at any given temperature, especially the distribution between typical planetary metal-core and silicate mantle components.  

\begin{figure}[ht]
 \centering
 \includegraphics[width=0.5\textwidth]{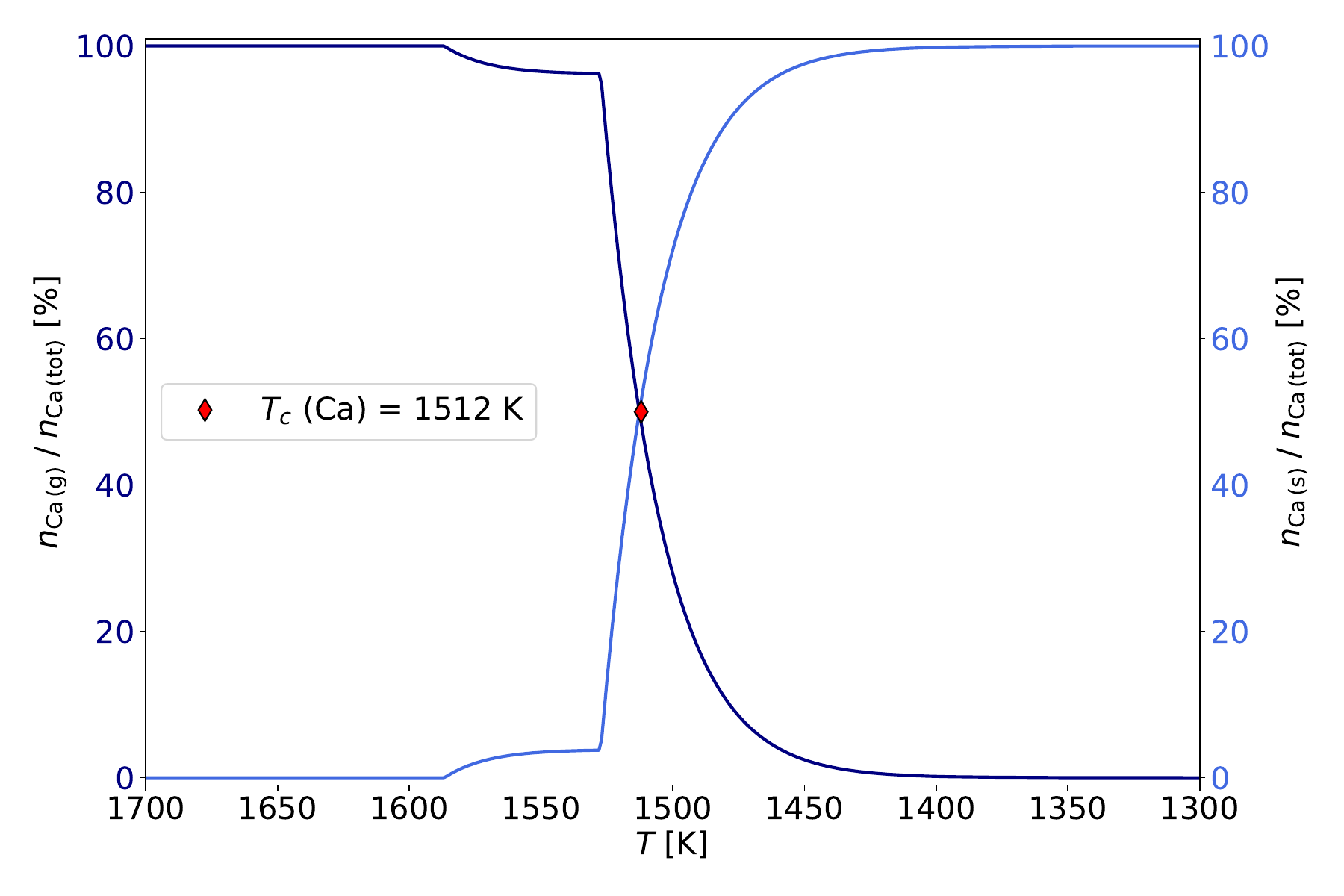}
 \caption{Example of the condensation curve and condensation temperature of a specific element, here \ce{Ca}. The dark blue curve shows the fraction of \ce{Ca} atoms bound in gas-phase species, the blue line shows the fraction of \ce{Ca} atoms bound in solid-phase species. The $T$-value of the intersection at 50\% of atoms in gas- and solid-phase signifies the 50\% condensation temperature.}
 \label{fig:condT_Ca}
\end{figure}

We implemented a computation of the condensation temperature of elements, as a common parameter to assess planet compositions. This is defined as the temperature at which 50\% of the total amount of an element is bound in solid-phase species. As an example, we show the condensation of \ce{Ca} in Fig. \ref{fig:condT_Ca}. The dark blue line denotes the fraction of \ce{Ca}-atoms bound in gas-phase species, $\left(\frac{n_{\ce{Ca(g)}}}{n_{\ce{Ca(tot)}}}\right)$, in \%, the light blue line the fraction bound in solid-phase species, $\left(\frac{n_{\ce{Ca(s)}}}{n_{\ce{Ca(tot)}}}\right)$. The curves intersect at $T = \SI{1512}{\K}$, signalling the 50\% condensation temperature of \ce{Ca}.

\begin{figure*}[ht]
  \begin{subfigure}[t]{0.5\textwidth}
   \centering
   \includegraphics[width=\textwidth]{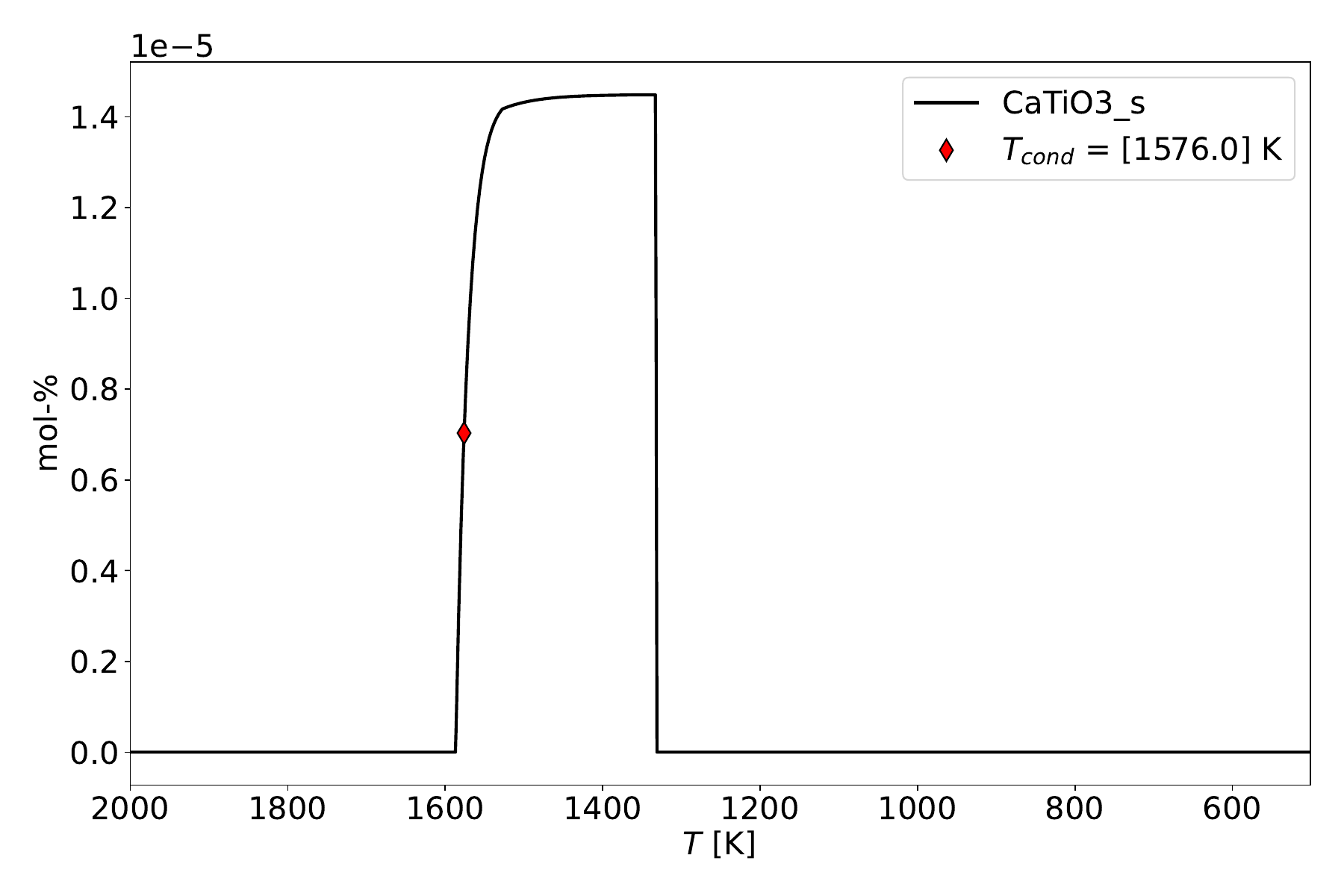}
  \end{subfigure}
  \hfill
  \begin{subfigure}[t]{0.5\textwidth}
   \centering
   \includegraphics[width=\textwidth]{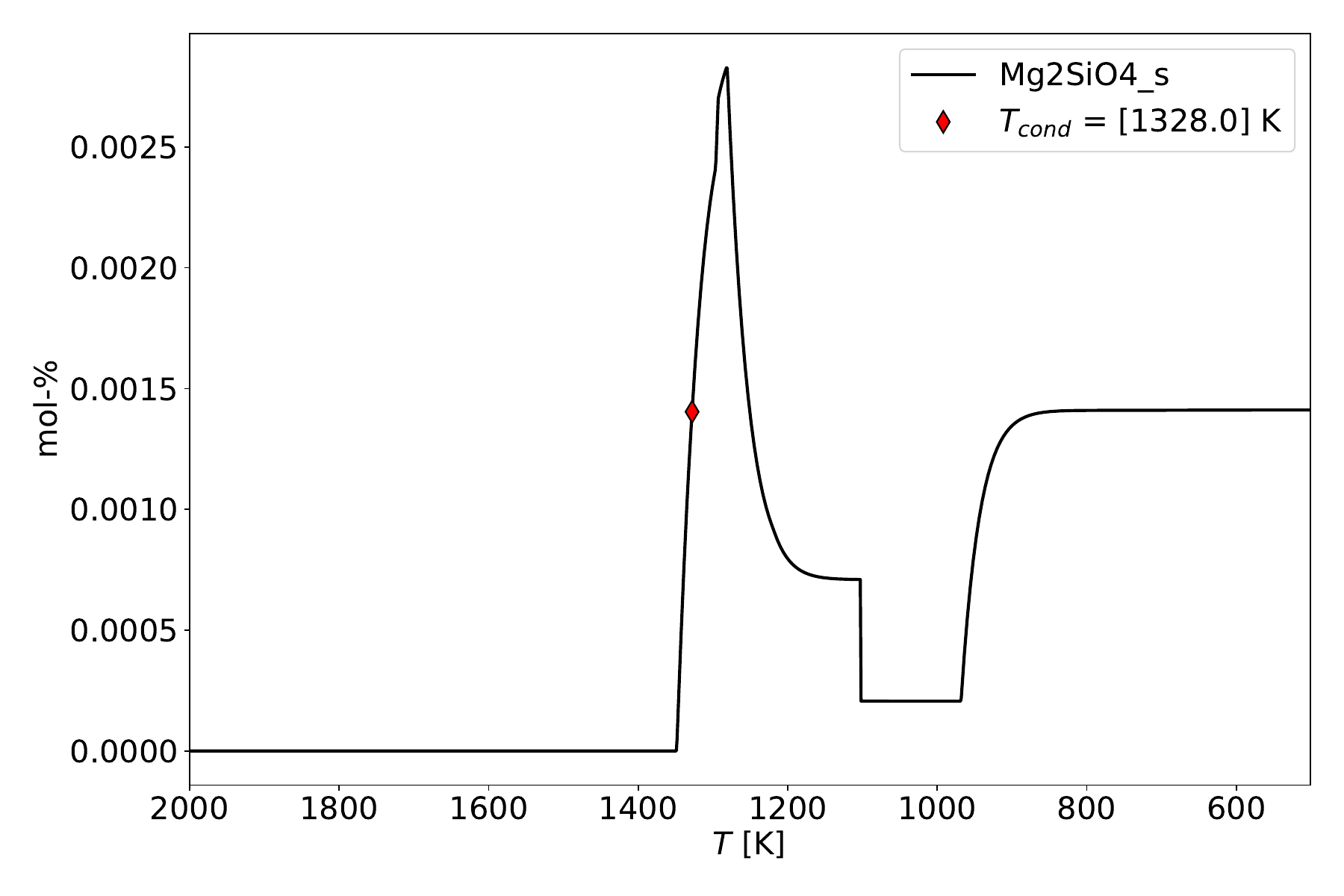}
  \end{subfigure}
  \caption{Examples of different types of condensation behaviours of solid species. We show the relative molar amount of the species present in the protoplanetary disk as a function of decreasing temperature to demonstrate our definition of 50\% condensation temperatures. Red diamonds show the computed 50\% condensation temperatures. \textit{Left panel}: Progression of perovskite \ce{CaTiO3(s)} (condensation and subsequent disintegration). \textit{Right panel}: Progression of forsterite \ce{Mg2SiO4(s)} (varying relative amounts).}
  \label{fig:cond_ex}
\end{figure*}

Additionally, we define a `condensation temperature' of specific solid-phase molecules (synonymously used for minerals) in order to analytically assess the appearance of condensates and sequences of phases containing specific elements. This value is, however, far less distinct than the condensation temperature of an element. There are two reasons for this: firstly, the maximum amount of a particular phase is not known a priori and depends on the temperature range considered; secondly, phases often disintegrate at lower temperatures in favour of more stable phases. This behaviour is exemplified in the left panel of Fig. \ref{fig:cond_ex} for perovskite (\ce{CaTiO3(s)}), which is superseded by \ce{Ti4O7(s)}. Some species show even more complex condensation behaviours, involving successions of increases and decreases in their relative amounts. This is demonstrated in the right panel of Fig. \ref{fig:cond_ex} for forsterite (\ce{Mg2SiO4(s)}), which is part of the intricate \ce{Mg}-\ce{Si}-chemistry in the protoplanetary disk. Irrespective of the precise curve shape, we only compute one condensation temperature per species, based on the 50\% level of the local maximum closest to the first onset of condensation. If a species is present in such small quantities that it cannot be distinguished from numerical noise, it is assumed to not have condensed in the simulation, and therefore no condensation temperature is reported. 

For the analysis of the composition of an exoplanet itself, we use the code's output of relative molar amounts of the chemical species that are stable in chemical equilibrium as a function of temperature. We convert this information into the relative amounts of the individual elements bound in solid species, using the structural formulae of the species. This procedure results in a temperature-series of the elemental composition of solid material in the disk. As default, we give the resulting composition in units of $\text{wt}-\%$, meaning the total mass of one element bound in solid species relative to the total mass of all elements in solid species, using the atomic weights data specified in Sect. \ref{sec:datatypes_sources}. To get the composition of an individual planet out of the temperature series, we have to specify its `formation temperature'. This temperature defines the point, at which the planetary material is taken out of the chemical equilibrium of the disk. This does not necessarily mean that the planet is fully formed at this temperature, nor that it corresponds to the final disk temperature at the location of the planet. It is sufficient that the planetesimals have become so large that their interiors are effectively shielded from the disk chemistry. For instance, the depletion pattern of volatile elements in the Earth suggests a formation temperature between
\SI{1100}{\K} and \SI{1400}{\K} \citep{Wang2019,Sossi2022}. We discuss our different models connecting the formation temperature to the planet's composition in Sect. \ref{sec:planet_comp_theory}.

%%%%%%%%%%%%%%%%%%%%%%%%%%%%%%%%%%%%%%%%%%%%%%%%%%%%%%%%%%%%%%%%%%%%%%%%%%%%%%%%%%%%%%%%%%%%%%%%%%%%%%%%%%%%%%%%%%%%%%%%%%%%%%%%%%%%%%%%%%%%%%%%%%%%%%%%%%%%%%%%%%%%%%%%%%%%%%%%%%%%%%%%%%%%%%%%%%%%%%
\section{Condensation temperatures and their dependence on pressure and element abundance}\label{sec:SimRes}
We used our code to look at condensation temperatures and their variability. The condensation temperatures of rocky species give a first idea of the building blocks of a planet, because only material that has condensed at the formation temperature of a planet can be accreted from the protoplanetary disk. The condensation temperature of elements takes this idea to a slightly more abstract level, as we are not looking at the specific material that can be accreted onto a planet anymore, but rather think about the final elemental composition of the planet, even after the original planetesimals have undergone chemical changes in the formation and consolidation of the planet, for instance due to thermal processes. This relies on the assumption that while the specific molecules that have been accreted might not be found in the final planet in their original form, their elemental proportions will be retained. Finally, the variability of the condensation temperatures gives an indication as to how applicable our understanding of the connection between the formation and composition of the Earth is to different planet formation regions in the disk and to exoplanetary systems.

To validate the results of our code externally, we compared them against benchmark results from the literature. In Sect. \ref{sec:SimRes_CT} we show our computed condensation temperatures of common rocky species (Sect. \ref{sec:SimRes_CT_rock}) and of elements (Sect. \ref{sec:SimRes_CT_element}) for a Solar System elemental abundance pattern at a constant pressure, and compare them against the results by \citet{Lodders2003a} and \citet{Wood2019}. In Sect. \ref{sec:SimRes_condT_var} we explore the variability of the condensation temperatures of elements as a function of the disk pressure and the stellar elemental abundance pattern.

\subsection{Condensation temperatures for a solar elemental abundance pattern and constant pressure}\label{sec:SimRes_CT}
First, we analyse the condensation temperatures of species and elements that can be derived for the solar elemental abundance pattern at the disk pressure associated with the formation of Earth. In order to also use our results as a benchmark test for our code, we used the same system parameters that were used in the studies we compare our results against, even if these values are not necessarily in line with the currently most accepted ones.

Namely, we used the Solar System elemental abundances pattern as reported by \citet{Lodders2003a} and a disk pressure of \SI{1e-4}{\bar}, which was found to represent the total pressure in the solar nebula near \SI{1}{AU} by \citet{Fegley2000}, and which was also used for the simulations of \citet{Lodders2003a} and \citet{Wood2019}. Our simulation covered a temperature range from \SI{2000}{\K} to \SI{300}{\K}, with a resolution of \SI{1}{\K}. The number of species included in our simulation (47 gases + 25 solids) was much smaller than in the comparison studies. The simulation parameters are summarised in Table \ref{tab:SimParams} in the Appendix.

We estimate very conservatively that different codes should return condensation temperatures of molecules within \SI{100}{\K} of each other and condensation temperatures of elements within \SI{20}{\K} for a given disk pressure and abundance pattern. This presumes that the most common elements and the majority of the most stable molecules are included in the simulation. This estimate is based on our experience regarding the response of our own code to variations in the molecule selection, the thermochemical data, and the definition of the condensation temperatures, in the case of molecular species. 

\subsubsection{Condensation temperatures of rocky species}\label{sec:SimRes_CT_rock}
For the validation of our simulated condensation temperatures of common rocky species, we used the benchmark results of the seminal work by \citet{Lodders2003a}. Their results have been computed with the \textsc{Condor}-code, which is not publicly available. Their thermochemical database contains 2000 gas-phase species and 1600 solids, which are all considered for the simulation. In their code, the chemical equilibration is based on equilibrium constants of formation, rather than Gibbs free energy, and the reported condensation temperatures pertain to the point at which the computed activity of a solid species reaches unity \citep{Lodders2003a}. Visually, this point would typically correspond to the onset of condensation, that is, the initial sharp change in gradient seen in our condensation curves of molecules (see Fig. \ref{fig:cond_ex} as an example). Depending on the slope of the curve, this temperature might easily be \SI{20}{\K} higher than our 50\% condensation temperature, as defined in Sect. \ref{sec:CompSol_Funcs_SimAnal}. 

Keeping this in mind, we found a high degree of agreement between the two sets of values, as shown in Table \ref{tab:Comp_condT_specs}. Our condensation temperatures are mostly within $\pm \SI{50}{\K}$ of the literature values. Our values are on average lower than those of \citep{Lodders2003a}, confirming our expectation based on the different definitions of the condensation temperature.

Regarding condensation sequences, we found that the order in which the molecules are expected to condense has been reproduced well with our code. The slight observed differences, as well as our failure to condense grossite (\ce{CaAl4O7}), are likely due to the fact that our code does not include solid solutions. The solid solutions of the olivine and pyroxene mineral groups especially will lead to a shift in the \ce{Mg} budget, potentially affecting the condensation of most of the shown species.

In conclusion, the found agreement between the two codes is much better than our conservative estimate. The combination of vagueness of the definition of the condensation temperature of a molecule and the large uncertainty in the thermochemical data itself (see Sect. \ref{sec:dataprocess}) suggests that one should not attach great meaning to their exact simulated values.

\begin{table}[ht]
\caption{Comparison of condensation temperatures of some common planetary species.}
 \centering
 \resizebox{0.5\textwidth}{!}{%
 \begin{tabular}{llm{11mm}m{12mm}p{10mm}}
 \hline\hline
   \multicolumn{2}{c}{} & \multicolumn{3}{c}{Condensation Temperature in [\si{\K}]}\\
   \multicolumn{2}{c}{Species} & \citet{Lodders2003a} & our code & deviation \\\hline
  \ce{Al2O3} & Corundum & 1677 & 1644 & $-33$\\
  \ce{CaTiO3} & Perovskite & 1593 & 1576 & $-18$\\
  \ce{CaAl4O7} & Grossite & 1542 & no cond. & n.a.\\
  \ce{Ca2Al2SiO7} & Gehlenite & 1529 & 1512 & $-17$\\
  \ce{MgAl2O4} & Spinel & 1397 & 1360 & $-37$\\
  \ce{CaAl2Si2O8} & Anorthite & 1387 & 1295 & $-92$\\
  \ce{Fe} & Iron & 1357 & 1331 & $-26$\\
  \ce{Mg2SiO4} & Forsterite & 1354 & 1328 & $-26$\\
  \ce{CaMgSi2O6} & Diopside & 1347 & 1359 & $+12$\\
  \ce{MgSiO3} & Enstatite & 1316 & 1262 & $-54$\\\hline
 \end{tabular}}
 \label{tab:Comp_condT_specs}
\end{table}

\subsubsection{Condensation temperatures of elements}\label{sec:SimRes_CT_element}

In contrast to the condensation of a rocky species, the 50\% condensation temperature of the elements (the temperature at which 50\% of the element is bound in solid species) is very sharply defined, since its maximum amount is known {a priori} from the given elemental abundance (cf. Sect. \ref{sec:CompSol_Funcs_SimAnal}). Additionally, the selection of species is less influential in the equilibrium computation, since the elements tend to only be exchanged between different solid-phase species after the initial onset of condensation, but do not to return to gas-phase species. Accordingly, not including a particular species does not affect the amount of the element being bound in solid-phase species, and consequently the 50\% condensation temperature.

Furthermore, the condensation temperatures of elements enable us to easily estimate the composition of a rocky planet. Most elements condense (10\% to 90\%) within a $T$-interval smaller than \SI{100}{\K}. This implies that there is hardly any of the element in solid form at temperatures above the 50\% condensation temperature, whereas all of it is in solid form at temperatures below. Hence, if the formation temperature of a planet is above the 50\% condensation temperature of an element, it will not contain significant amounts of this element. Otherwise, the element will have a similar relative abundance in the planet as in its host star. If the formation temperature of the planet is close to the condensation temperature of an element, this element will likely be depleted to some degree in the planet.

For the comparison, we again used \citet{Lodders2003a}, as well as the more recent study of \citet{Wood2019}. The latter applied the \textsc{PHEQ} code, developed by and described in \citet{Wood1993}, which is very similar in its thermochemical approach to our code.

\begin{table}[ht]
\caption{Comparison of 50\% condensation temperatures of major rock-forming elements.}
 \centering
 \resizebox{0.5\textwidth}{!}{%
 \begin{tabular}{llll}
 \hline\hline
   & \multicolumn{3}{l}{Condensation Temperature in \si{\K}}\\
  Element & \citet{Lodders2003a} & \citet{Wood2019} & our code \\\hline
  \ce{Al} & 1653 & 1652 & 1643 \\
  \ce{Ti} & 1582 & 1565 & 1575 \\
  \ce{Ca} & 1517 & 1535 & 1512 \\
  \ce{Mg} & 1336 & 1343 & 1331 \\
  \ce{Fe} & 1334 & 1338 & 1329 \\
  \ce{Si} & 1310 & 1314 & 1305 \\\hline
 \end{tabular}}
 \label{tab:Comp_condT_el}
\end{table}
\begin{figure}[ht]
 \centering
 \includegraphics[width=0.5\textwidth]{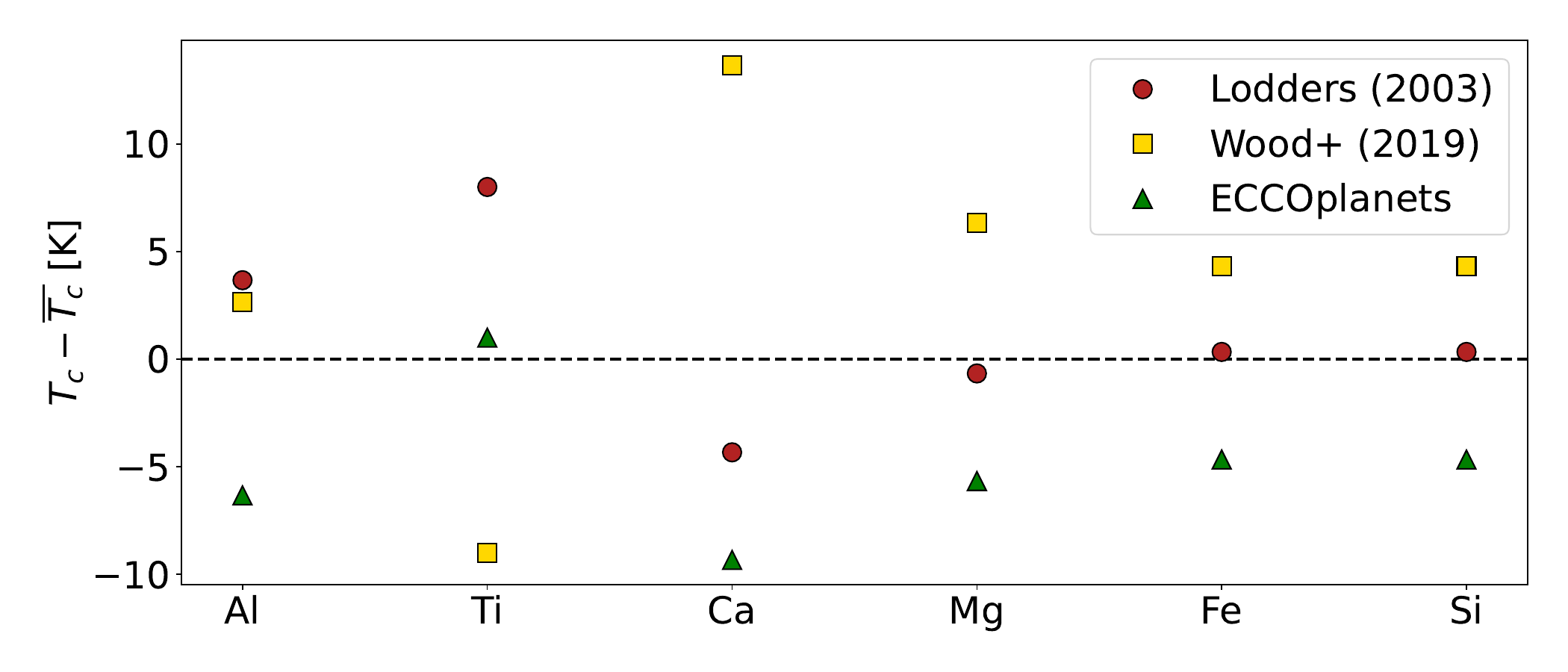}
 \caption{Graphical comparison of deviation of 50\% condensation temperatures of the major rock-forming elements, minus the mean over the three values for each element.}
 \label{fig:Comp_condT_el}
\end{figure}

The agreement with \citet{Lodders2003a} and \citet{Wood2019} for the condensation temperatures of major rock-forming elements are excellent, as shown in Table \ref{tab:Comp_condT_el}. The average discrepancy from the mean condensation temperature of each element is below \SI{5}{\K} and the largest deviation is below \SI{15}{\K} (cf. Fig. \ref{fig:Comp_condT_el}).

We conclude that the 50\% condensation temperatures of the most common planet-building elements are not sensitive to details of the used condensation algorithm. In other words, even our very simplistic approach with only a few included species, will return results with a projected error below $\pm\SI{10}{\K}$, for a given elemental abundance pattern and disk pressure. 

\subsection{Dependence of the condensation temperatures of elements on system parameters}\label{sec:SimRes_condT_var}
We explored the variability of the condensation temperature of elements as a function of disk pressure and elemental abundance pattern. While it has been widely acknowledged that the chemical processes in the protoplanetary disk are controlled by its elemental composition and especially its \ce{C}/\ce{O} and \ce{Mg}/\ce{Si} ratio \citep{Bond2010a,Carter-Bond2012,Thiabaud2014,Moriarty2014,Dorn2019,Bitsch2020}, the condensation temperatures of the elements are sometimes implicitly treated almost as material constants, at least within certain limits of system parameters \citep[see e.g.][]{Wang2019,Wang2019b,Spaargaren2020}. 

In Sect. \ref{sec:condT_var_p} we explore the influence of the disk pressure, and in Sect. \ref{sec:condT_var_ab} we analyse the effect of a variation in the elemental abundance pattern. In Sect. \ref{sec:condT_var_imp} we briefly discuss implication of our findings.

\subsubsection{Dependence on disk pressure}\label{sec:condT_var_p}
The condensation temperatures of elements are often reported for a total disk pressure of \SI{1e-4}{\bar}. In general, however, the disk pressure depends both on the radial distance from the central star, the vertical distance from the mid-plane, and the total material within the system \citep{Fegley2000}. 

\begin{figure*}
 \centering
 \includegraphics[width=0.7\textwidth]{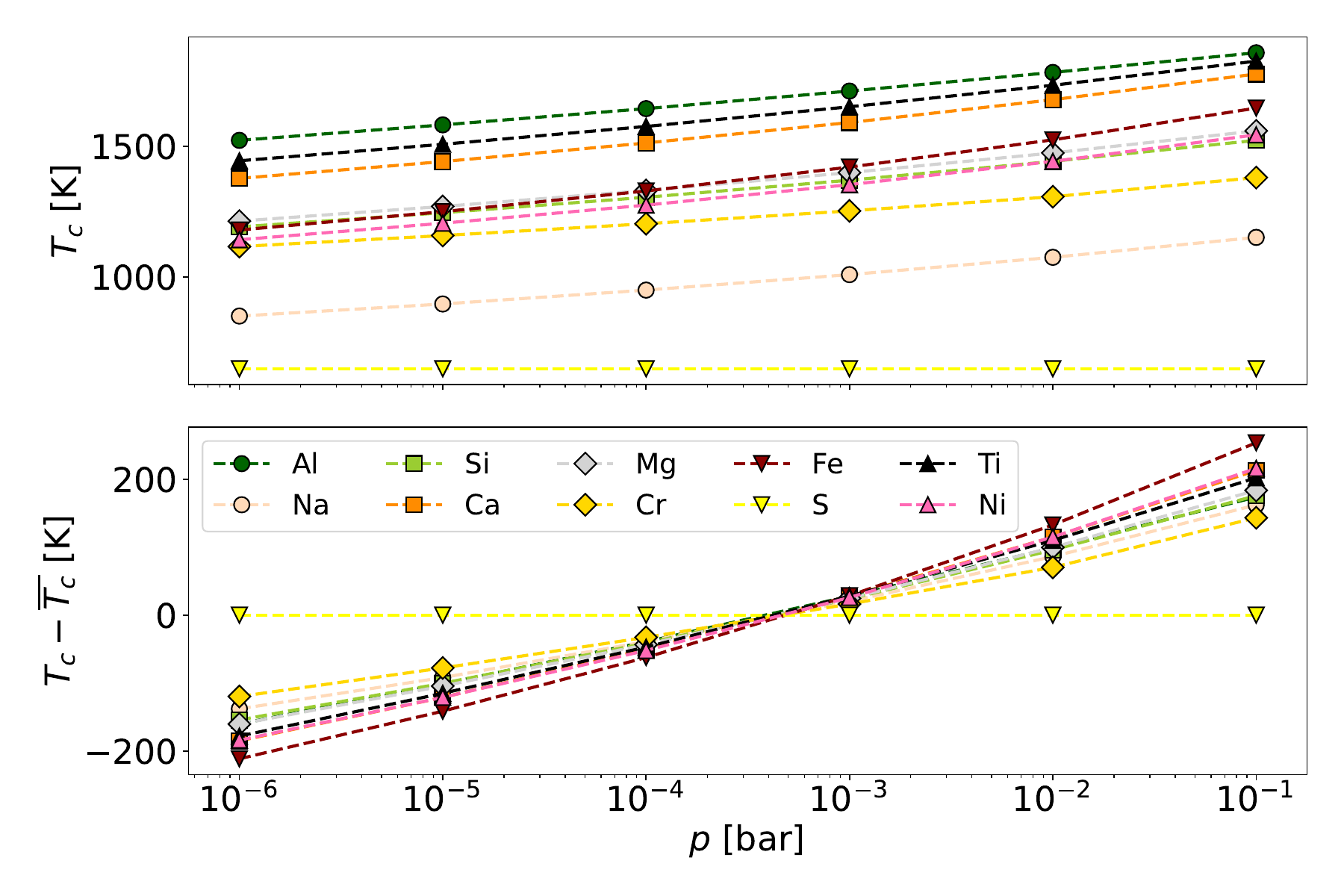}
 \caption{Dependence of the 50\% condensation temperature ($T_c$) of elements on the disk pressure. Markers represent simulated condensation temperatures, and corresponding dashed lines are only to guide the eye. Colours, as denoted in the legend, are the same for both panels. All simulations use the solar elemental abundance pattern recommended by \citet{Lodders2003a}. \textit{Top panel}: 50\% condensation temperature of major planet-building elements as a function of different disk pressures. \textit{Bottom panel}: Deviation of the 50\% condensation temperature from the respective mean condensation temperature.}
 \label{fig:condTp}
\end{figure*}

We varied the disk pressure logarithmically between \SI{1e-6}{\bar} and \SI{1e-1}{\bar}, while keeping all other parameters of the simulation constant. Depending on the disk model, this range of pressure values might correspond to a radial distance range from \SI{0.1}{AU} to \SI{5}{AU} in a Solar-System-like disk \citep{Fegley2000}, that is, distances within the water snow line, where we expect to find rocky planets. 

As shown in the top panel of Fig. \ref{fig:condTp}, we found an overall trend of a higher disk pressure corresponding to a higher condensation temperature of the elements. Raising the pressure in a system where nothing else is changed, is equivalent to increasing the particle concentration. This implies an increase in the reaction rates. Since the pressure raises the effective concentration of all species equally, we found a quantitatively very similar relation between the disk pressure and the condensation temperature for all analysed elements, except for \ce{S}. 

This can be seen particularly clearly in the bottom panel of Fig. \ref{fig:condTp}, where we show the deviation of the condensation temperature of an element at a given disk pressure from the element's mean condensation temperature within the analysed pressure range. For the analysed pressure range all species have their mean condensation temperature at roughly the same pressure (between \SI{4e-4}{\bar} and \SI{5e-4}{\bar}). Also, the deviation from this mean is similar for all elements at each disk pressure. On average, the condensation temperature of all elements except for \ce{S} increases by \SI{357\pm57}{\K} over the analysed five orders of magnitude in disk pressure. \ce{S} behaves differently, its condensation temperature does not change at all over the pressure range.\footnote{\label{fn:S-O}The deviation is, however, not relevant for our argument, because the condensation temperature of \ce{S} is much lower than those of the other main rock-forming elements.}

The consistency in pressure versus condensation temperature relations between the different elements has implications for the analysis of planet formation at different radial locations within the protoplanetary disk. In disk models, both the midplane temperature and the disk pressure decrease with increasing distance from the central star. Generally, when simulating planet formation, both pressure and temperature are considered in combination. However, since the disk pressure affects the condensation temperatures of all elements very similarly, the variation in disk pressure does not change the equilibrium chemistry and equilibrium composition as a function of temperature qualitatively, but only shifts the equilibrium composition to higher temperatures (for an increased pressure) or to lower temperatures (for a decreased pressure). The small variations in the pressure response of the elements' condensation temperature will likely be rendered insignificant by the fact that a planet does not comprise material of one specific ($T$-$p$)-equilibrium condition but rather a mixture over a range of conditions. As shown in Fig. \ref{fig:condTp}, the greater the change in pressure, the larger the difference in $T_c$ between the elements. Our argument is therefore only valid for small pressure ranges.

\subsubsection{Dependence on the elemental abundance pattern}\label{sec:condT_var_ab}
There have been many studies assessing the diversity of exoplanetary compositions as a result of the changed equilibrium chemistry in a protoplanetary disk, due to variations in its elemental abundance pattern \citep[e.g.][]{Bond2010a,Carter-Bond2012,Thiabaud2014,Moriarty2014,Dorn2019,Bitsch2020,Jorge2022}. Certain element ratios control which molecular species will form out of the available elements. Different molecules can have vastly different condensation temperatures even if they consist of similar elements. The species in which an element is predominantly bound, generally determines its 50\% condensation temperature. 

To systematically analyse the influence of the elemental abundance pattern on the condensation temperatures of the elements, we ran condensation simulations for synthetic abundance patterns, only varying one key element ratio at a time. We explored the role of the overall metallicity and the element ratios \ce{C}/\ce{O}, \ce{Mg}/\ce{Si}, \ce{Fe}/\ce{O}, and \ce{Al}/\ce{Ca}. We specify metallicities logarithmically and normalised to solar values: 
\begin{equation}
\left[\ce{M}/\ce{H} \right] = \log\left(\frac{N_M}{N_H}\right)_{\rm{star}} - \log\left(\frac{N_M}{N_H}\right)_{\rm{sun}},
\end{equation}
where $N_M$ is the sum of the relative number of atoms in the system of all elements larger than \ce{He}, and $N_H$ the relative number of \ce{H} atoms. All other element ratios are given as non-normalised number ratios, for instance, `\ce{C}/\ce{O}' means
\begin{equation}
\ce{C}/\ce{O} = \left(\frac{N_C}{N_O}\right)_{\rm{star}}.
\end{equation}

\begin{figure*}
 \centering
 \includegraphics[width=1\textwidth]{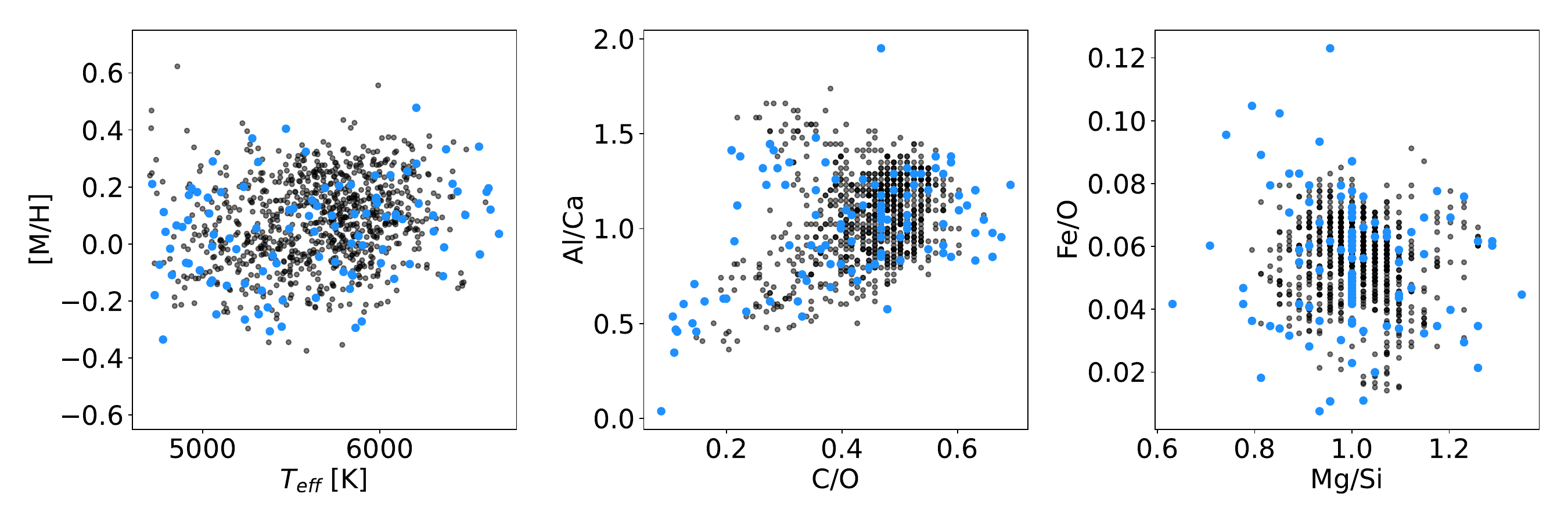}
 \caption{Parameter range of the \citet{Brewer2016} stellar database. Grey background markers represent all stars with $\log g > 3.5$ and $ S/N > 100$ of the database, and light blue foreground markers represent the stellar sample studied here. \textit{Left Panel}: Effective temperature versus metallicity. \textit{Middle Panel}: \ce{C}/\ce{O} ratio versus \ce{Al}/\ce{Ca} ratio. \textit{Right Panel}: \ce{Mg}/\ce{Si} ratio versus \ce{Fe}/\ce{O} ratio.}
 \label{fig:Brewer_stars}
\end{figure*}

As a basis for our analysis, we used the \citet{Brewer2016} catalogue of $1617$ F, G, and K stars. To ensure the reliability of the abundance data, we only took stars into account whose spectra have a signal-to-noise ratio (S/N) larger than $100$. Furthermore, to avoid giant stars, we excluded all stars with $\log g \leq 3.5$ \cite[compare][]{Harrison2018}. We used the remaining $964$ stars to (1) generate a representative abundance pattern as a starting point for the element ratio variations, (2) determine the parameter ranges of the element ratios we were interested in, and (3) pick roughly $100$ stars, covering the whole parameter space, to verify that any trends found in the analysis of the synthetic data are also followed by the real stellar data. Figure \ref{fig:Brewer_stars} shows the parameter ranges ($T_{\rm{eff}}$ versus metallicity, \ce{C}/\ce{O} versus \ce{Al}/\ce{Ca}, and \ce{Mg}/\ce{Si} versus \ce{Fe}/\ce{O}) covered by the \citet{Brewer2016} stars, and the distribution of the sample of comparison stars.

\begin{figure}[ht]
 \centering
 \includegraphics[width=0.5\textwidth]{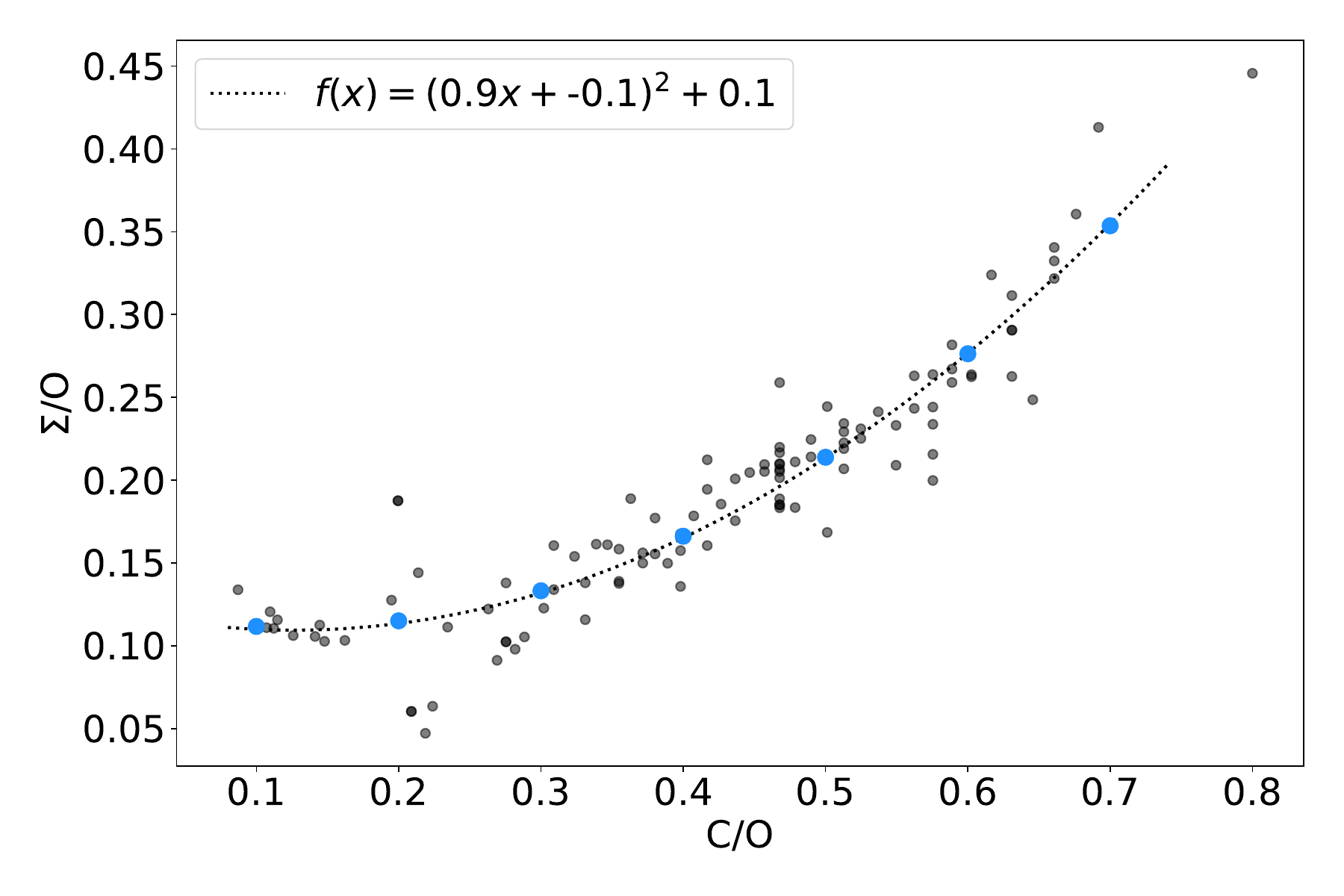}
 \caption{Correlation between the \ce{C}/\ce{O} and $\Sigma$/\ce{O}, where $\Sigma = N_{\ce{Mg}}+N_{\ce{Si}}+N_{\ce{Fe}}+N_{\ce{Ca}}+N_{\ce{Al}}$. Grey background markers represent the studied stellar sample, and the dotted line shows the fit quadratic fit to the data. The light blue foreground markers show the synthetic data for the \ce{C}/\ce{O} analysis.}
 \label{fig:CO_art_func}
\end{figure}

For most variations of the element ratios, we kept the abundance of one element constant in our representative abundance pattern, and only varied the other. For the overall metallicity, only the \ce{H} abundance was changed. For the \ce{Al}/\ce{Ca}, \ce{Mg}/\ce{Si}, and \ce{Fe}/\ce{O} ratios, \ce{Al}, \ce{Mg}, and \ce{Fe} were varied, respectively. The \ce{C}/\ce{O} ratio was treated differently. In the studied stellar sample, there is a strong correlation between the \ce{C}/\ce{O} ratio and the ratio of \ce{O} to the sum of other abundant elements, such as \ce{Mg}, \ce{Si}, and \ce{Fe}. In order to avoid a distortion of the analysis due to an unrealistic abundance pattern of the synthetic data, we approximated this correlation with a parabola, as shown in Fig. \ref{fig:CO_art_func}, and adapted both the \ce{C} and \ce{O} abundance accordingly.    

\begin{table}[]
\caption{Summary of the influence of the variation in different element ratios on the condensation temperature of elements.}
    \centering
    \begin{tabular}{ll|p{20mm}p{30mm}}
    \hline\hline
        \multicolumn{2}{l|}{} & \multicolumn{2}{c}{curve shape}\\
        \multicolumn{2}{l|}{} & log-linear & amorphous \\\hline
         \multirow{2}{*}{\rotatebox[origin=c]{90}{affects  }}
        & many elements & $\left[\ce{M}/\ce{H} \right]$ & \ce{C}/\ce{O} \\
        & few elements & \ce{Al}/\ce{Ca} (for \ce{Al}) \newline \ce{Fe}/\ce{O} (for \ce{Fe}) & \ce{Al}/\ce{Ca} (for \ce{Ca}) \newline \ce{Mg}/\ce{Si} (for \ce{Mg} \& \ce{Si})\\\hline
    \end{tabular}
    
    \label{tab:summary_el-rat}
\end{table}

Our tests show that an element ratio can affect the condensation temperatures in different ways. We can differentiate the effects in terms of the number of affected elements, and the curve shape of the correlation between the ratio and condensation temperature. Table \ref{tab:summary_el-rat} shows the summary of our findings. The overall metallicity and \ce{C}/\ce{O} ratio have the most profound impact on the condensation temperatures. These two variations stand out, because (1) they affect a large number of elements, (2) the magnitude of change in condensation temperatures is high, and (3) the correlation between the ratio and the condensation temperature of all affected elements is systematic. We therefore limit our discussion to these two parameters and only cover the others in a cursory fashion.

\begin{figure*}
 \centering
 \includegraphics[width=0.7\textwidth]{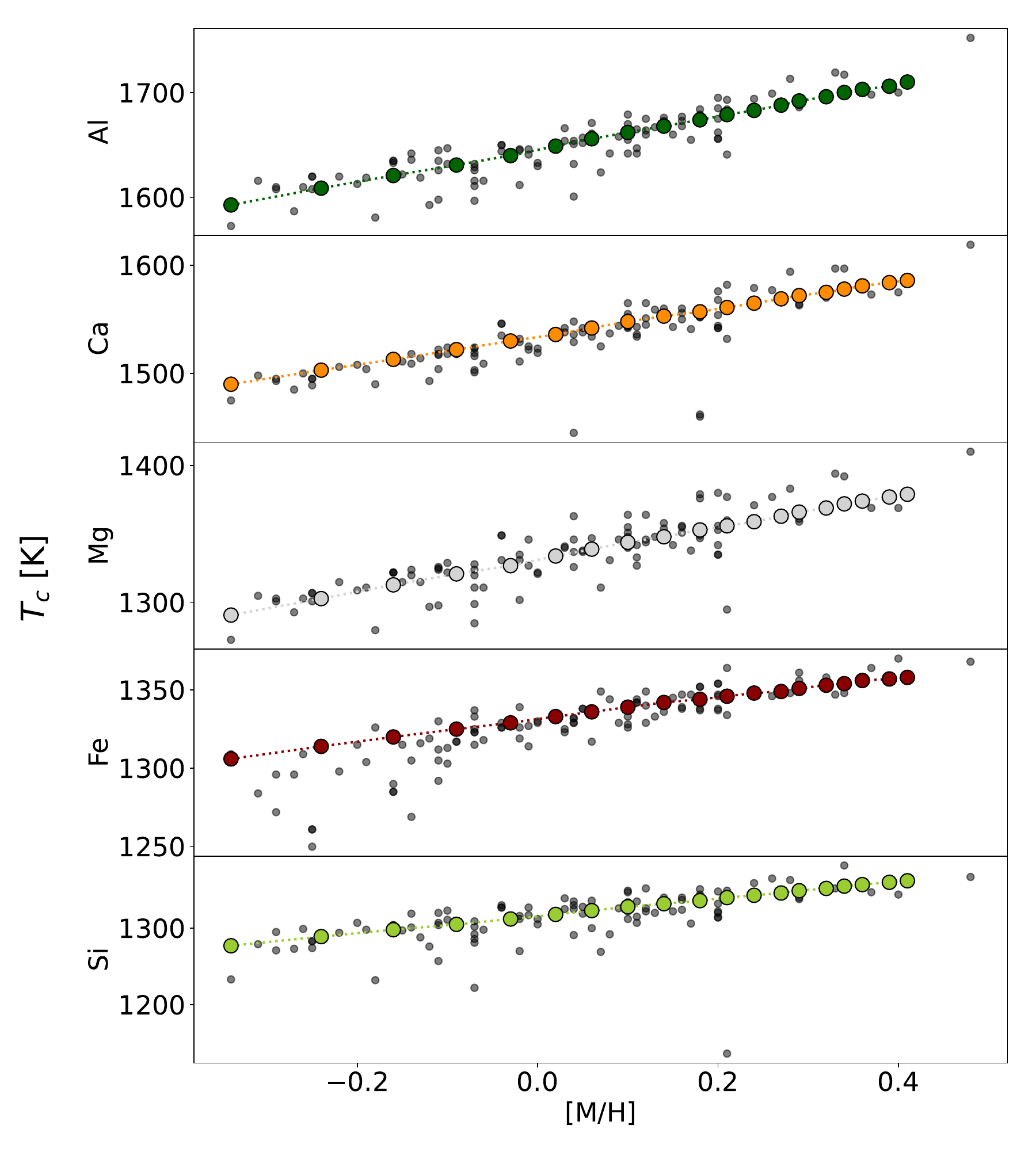}
 \caption{Correlation between the overall metallicity of the system and the condensation temperature of some major planet-building elements. The coloured circles in the foreground show the simulation results of the synthetic abundance patterns, and the grey circles in the background show the simulation results of a representative subset of approximately 100 stars from the \citet{Brewer2016} database. All simulations were run at a constant disk pressure of $p=10^{-4}~\si{bar}$.}
 \label{fig:MH_condT}
\end{figure*}

In Fig. \ref{fig:MH_condT}, we show the influence of the overall metallicity of the system on the condensation temperatures of a selection of common elements. The coloured foreground markers represent the simulation result for the synthetically varied metallicity, the grey background markers show the simulation results of the random sample of comparison stars. The figure clearly demonstrates a linear correlation between the logarithmic metallicity and condensation temperature for all elements. We found an increase in condensation temperature between \SI{52}{\K} (\ce{Fe}) and \SI{117}{\K} (\ce{Al}) over the covered metallicity range. Despite the fact that the abundance patterns of the comparison stars are quite diverse (cf. Fig. \ref{fig:Brewer_stars}), the log-linear correlation between metallicity and condensation temperatures can also be found there.\footnote{The deviation of the \ce{Fe} condensation temperatures at low metallicities is caused by the superposition of the effect of disproportionately low \ce{Fe}-abundances in the stellar sample \citep[`$\alpha$-enhancement'; see e.g.][]{Gebek2022}.} The median deviation of the condensation temperatures from the interpolation curve of the synthetic simulation results is below \SI{20}{\K} for all the elements.

The effect of the overall metallicity is reminiscent of the effect of disk pressure variations, seen in Sect. \ref{sec:condT_var_p}. This is unsurprising as both variations effectively change the relative number of rock-building particles per volume, that is, the chemical reaction rates. 

We have found a similar log-linear correlation for the \ce{Fe}/\ce{O} ratio, which does, however, only affect the condensation temperature of \ce{Fe} itself. Again, we suspect this to be explainable by the fact that we increased the partial pressure of \ce{Fe}. Since its dominant solid species in our simulation is pure solid iron, this increased partial pressure does not shift the balance of any other chemical reaction. The result would apply analogously to similar condensation patterns, where the dominant solid species do not include any other elements, such as the \ce{Ni} condensation.

\begin{figure*}
 \centering
 \includegraphics[width=0.7\textwidth]{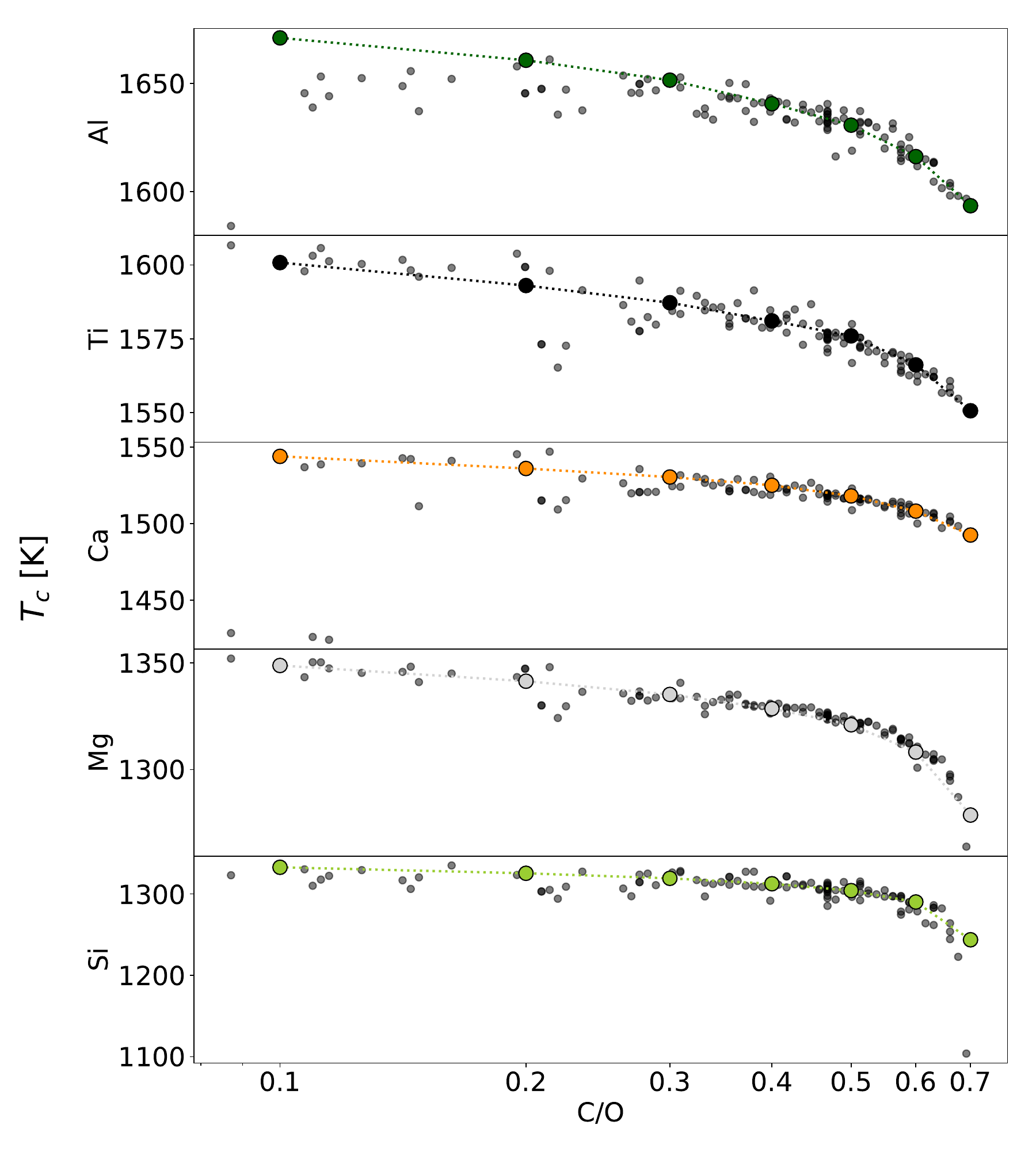}
 \caption{Correlation between the \ce{C}/\ce{O} ratio and the condensation temperature of some major planet-building elements, corrected for the metallicity effect. The coloured circles in the foreground show the simulation results of the synthetic abundance patterns, and the grey circles in the background show the simulation results of a representative subset of approximately 100 stars from the \citet{Brewer2016} database. All simulations were run at a constant disk pressure of $p=10^{-4}~\si{bar}$.}
 \label{fig:CO_condT}
\end{figure*}

In Fig. \ref{fig:CO_condT}, we show the influence of the \ce{C}/\ce{O} ratio on the condensation temperature of some common elements. Again, the synthetic simulation results are depicted with coloured foreground markers, the comparison stars with grey background markers. It is important to note that in this figure the effect of the metallicity, as described above, has already been removed from the results. The order of magnitude of the change in condensation temperature caused by the variation of the \ce{C}/\ce{O} ratio is similar to that of the metallicity. There are, however, also several qualitative differences. The most obvious difference is that an increase in the \ce{C}/\ce{O} ratio causes a decrease in condensation temperatures, in contrast to the increase caused by a higher metallicity. Also, while there certainly seems to be a systematic effect of the \ce{C}/\ce{O} ratio on all condensation temperatures, the correlation is not log-linear. Finally, while the metallicity affected all condensation temperatures, the \ce{C}/\ce{O} ratio only affects elements whose dominant species contain \ce{O}, for instance, the condensation of \ce{Fe} and \ce{Ni} are unaffected. 

\begin{figure*}[ht]
  \begin{subfigure}[t]{0.5\textwidth}
   \centering
   \includegraphics[width=\textwidth]{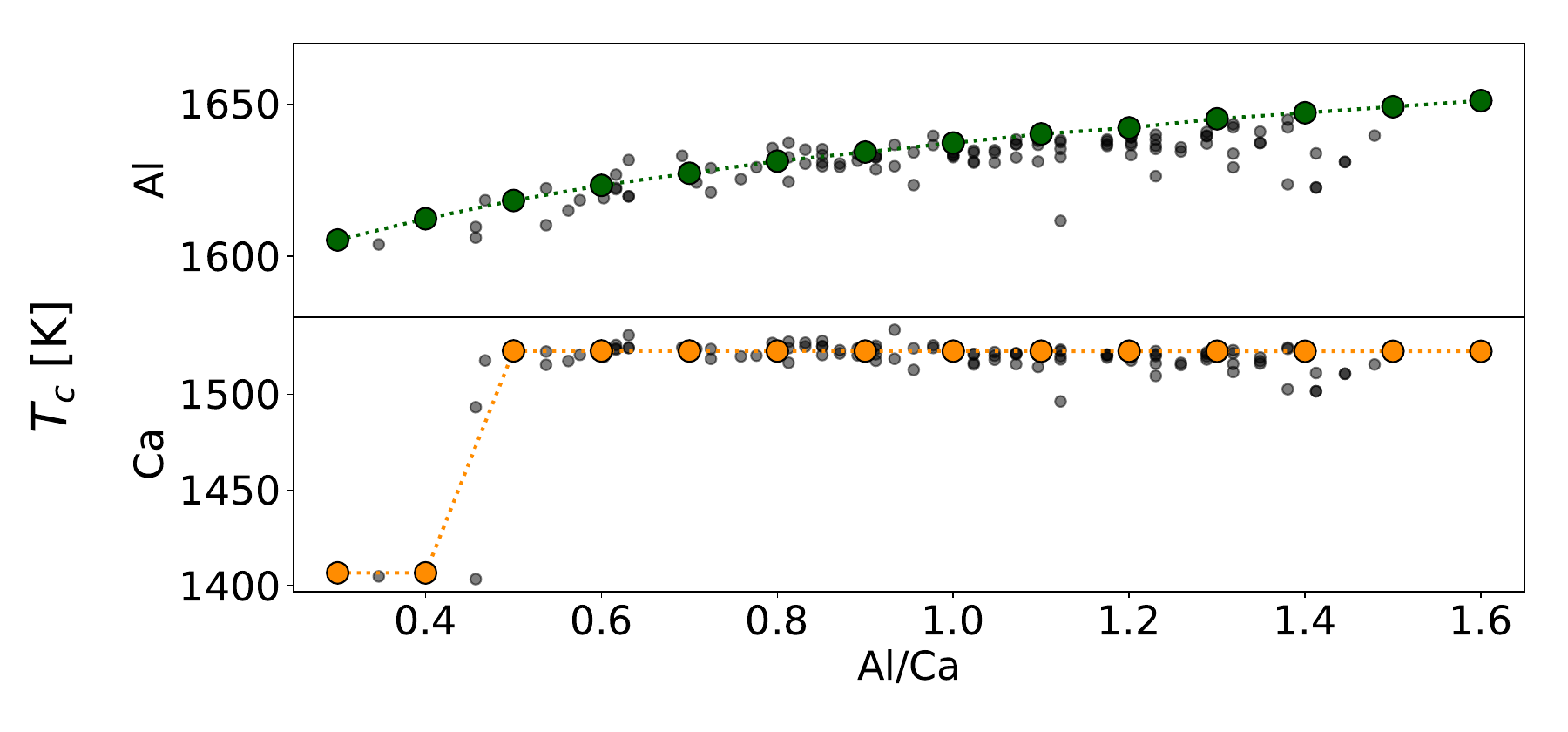}
  \end{subfigure}
  \hfill
  \begin{subfigure}[t]{0.5\textwidth}
   \centering
   \includegraphics[width=\textwidth]{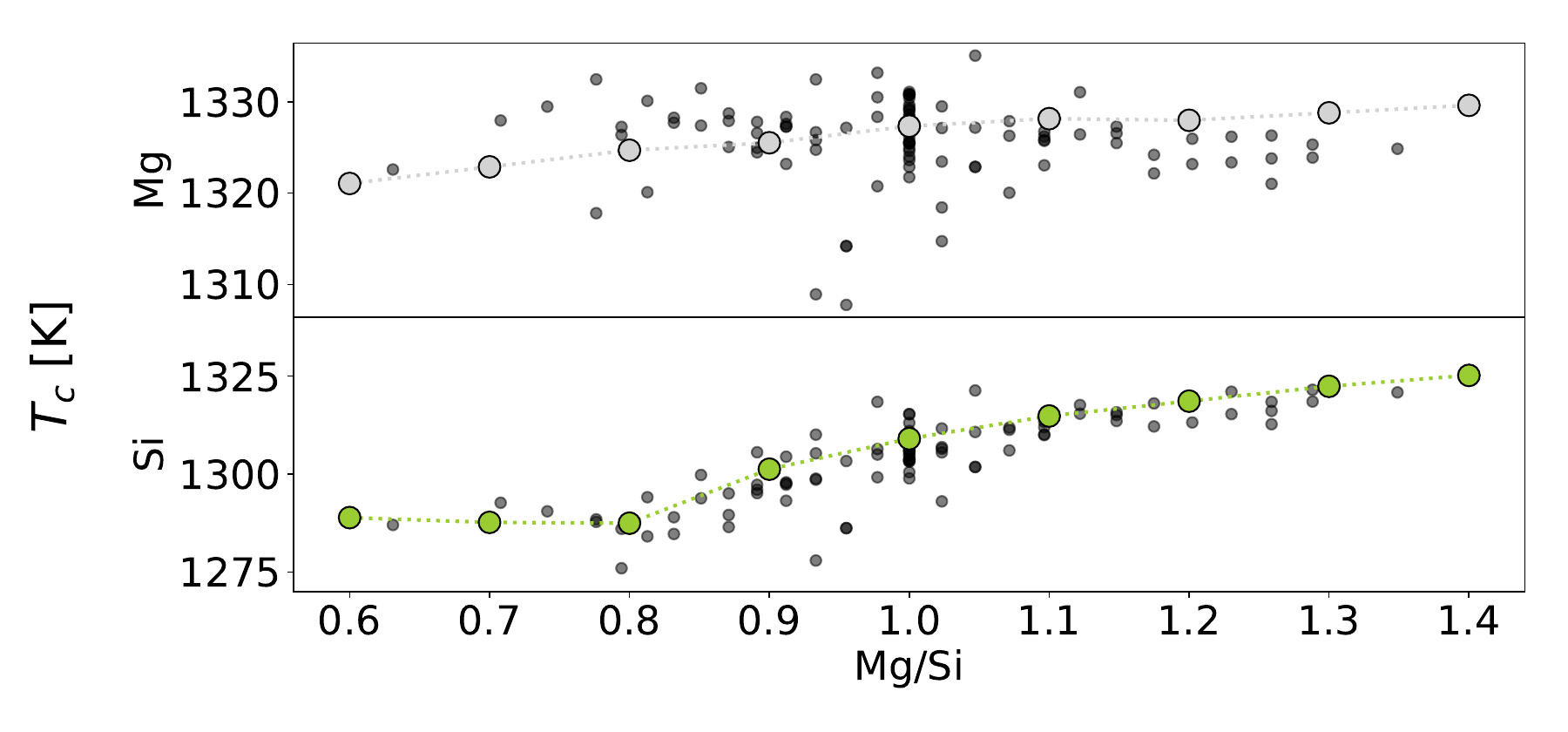}
  \end{subfigure}
  \caption{Correlation between element ratios and the condensation temperatures, corrected for the metallicity and \ce{C}/\ce{O} effect. The coloured circles in the foreground show the simulation results of the synthetic abundance patterns, and the grey circles in the background show the simulation results of a representative subset of approximately 100 stars from the \citet{Brewer2016} database. All simulations were run at a constant disk pressure of $p=10^{-4}~\si{bar}$. \textit{Left panel}: Influence of the \ce{Al}/\ce{Ca} ratio on the condensation temperature of \ce{Al} and \ce{Ca}. \textit{Right panel}: Influence of the \ce{Mg}/\ce{Si} ratio on the condensation temperature of \ce{Mg} and \ce{Si}.}
  \label{fig:other-rats_condT}
\end{figure*}

The expected correlation mapped out by the synthetic data is followed exceptionally well by the real data simulations, especially for the range $0.3 \leq \ce{C}/\ce{O} \leq 0.7$. For the whole parameter range, the median deviation from the expected curve is below \SI{10}{\K} for all elements. For \ce{C}/\ce{O} values below $0.3$, the real data condensation temperatures of \ce{Al} and \ce{Ca} do not follow the synthetic data well. This is caused by the superposition of the influence of the \ce{Al}/\ce{Ca} ratio. The diverging systems coincidentally all feature a particularly low \ce{Al}/\ce{Ca} ratio. Our tests with synthetically varied \ce{Al}/\ce{Ca} ratios have shown that it causes a roughly log-linear increase of the \ce{Al} condensation temperatures and a step-function increase for \ce{Ca} (see Fig. \ref{fig:other-rats_condT}, left panel). This explains why the condensation temperatures of \ce{Al} shown in Fig. \ref{fig:CO_condT} gradually taper off from the expectation curve, whereas the \ce{Ca} condensation temperatures suddenly jump to values more than \SI{100}{\K} below the expectation.

These differences between the effect of the variations in metallicity and in the \ce{C}/\ce{O} ratio allude to different underlying mechanisms. We have argued that an increase in metallicity implies increasing the number of all reactants per volume, thereby increasing all reaction rates. In contrast, changing the \ce{C}/\ce{O} ratio tilts the reaction balances in the formation of many major species, by only changing the availability of one of the reactants or by changing them to different degrees. The effect of a changed reaction balance is far more difficult to predict than the effect of globally increased reaction rates. Reaction balances are particularly strongly affected when the involved element ratios in a system are typically close to unity, because then a change in the ratio can imply that the availability one of the reactants is exhausted before the other, inhibiting this reaction. This is the case for the \ce{C}/\ce{O}, \ce{Mg}/\ce{Si}, and \ce{Al}/\ce{Ca} ratios.

\subsubsection{Implications of the variability in condensation temperatures}\label{sec:condT_var_imp}

Our findings regarding the variation of the condensation temperatures of elements have several implications. Most importantly, a combination of variations in pressure and elemental abundance pattern, even over a moderate parameter range, can easily change the condensation temperatures of elements by more than \SI{100}{\K}. This needs to be taken into account, when they are used to estimate planet compositions in other stellar system, for instance when applying the elemental devolatilization pattern of the Earth to exoplanets \citep{Wang2019b,Spaargaren2022} or in the context of white dwarf pollution \citep{Jura2014,Farihi2016,Harrison2018,Wilson2019,Bonsor2020,Veras2021,Xu2021}. 

We have, however, seen that the overall metallicity and the disk pressure affect all elements very similarly. Neglecting those will likely not be of great consequence to any derived exoplanetary compositions. That is to say, the computed element ratios would agree well with a model taking these parameters into account over the whole simulation range, but these ratios would be predicted for shifted radial distances. 

Other variations in the elemental abundance pattern, however, cause more unpredictable changes to the condensation temperatures of some elements. As a result, both the sequence in which the elements condense, as well as the difference in their condensation temperatures can be significantly altered. These changes entail substantial qualitative deviations in the most likely composition of a planet expected to form in a given system compared to its Solar System analogue. We explore this point further in the next section (Sect. \ref{sec:SimRes_PComp}).

Furthermore, our findings give us an idea of the potential influence of the uncertainty of stellar abundance measurements. While the uncertainties of the abundances of most planet-building elements might be lower than $\pm\SI{0.03}{dex}$ for many well-studied F, G, and K stars \citep{Brewer2016}, and even an uncertainty at the $\pm\SI{0.01}{dex}$ level seems feasible for these stars \citep{Bedell2014}, the situation is generally much worse. For M-dwarfs, where abundance measurements are in their infancy, typical errors exceed $\pm\SI{0.1}{dex}$ \citep{Souto2017}. As we see in Fig. \ref{fig:CO_condT}, a difference of $\pm\SI{0.1}{dex}$ in the \ce{C}/\ce{O} ratio can signify a difference of some tens of Kelvins in the condensation temperature of certain elements, at least at the upper end of our tested range, that is, $\ce{C}/\ce{O} \geq 0.5$. An in-depth analysis of the impact of uncertainty is available in \citet{Hinkel2018}.

\section{Exoplanet compositions}\label{sec:SimRes_PComp}
We now look at the bulk composition of rocky planets around chemically different stars. To emulate the dynamical formation of planets we compare different methods of assembling the solid material from our equilibrium condensation simulation. To externally validate our results and qualitatively assess the merits of our composition models, we compare them against the $n$-body simulations of \citet{Bond2010a} (in this Sect. abbreviated as B10).

\subsection{Derivation of planet compositions}\label{sec:planet_comp_theory}
\begin{figure}[ht]
    \centering
    \includegraphics[width=0.5\textwidth]{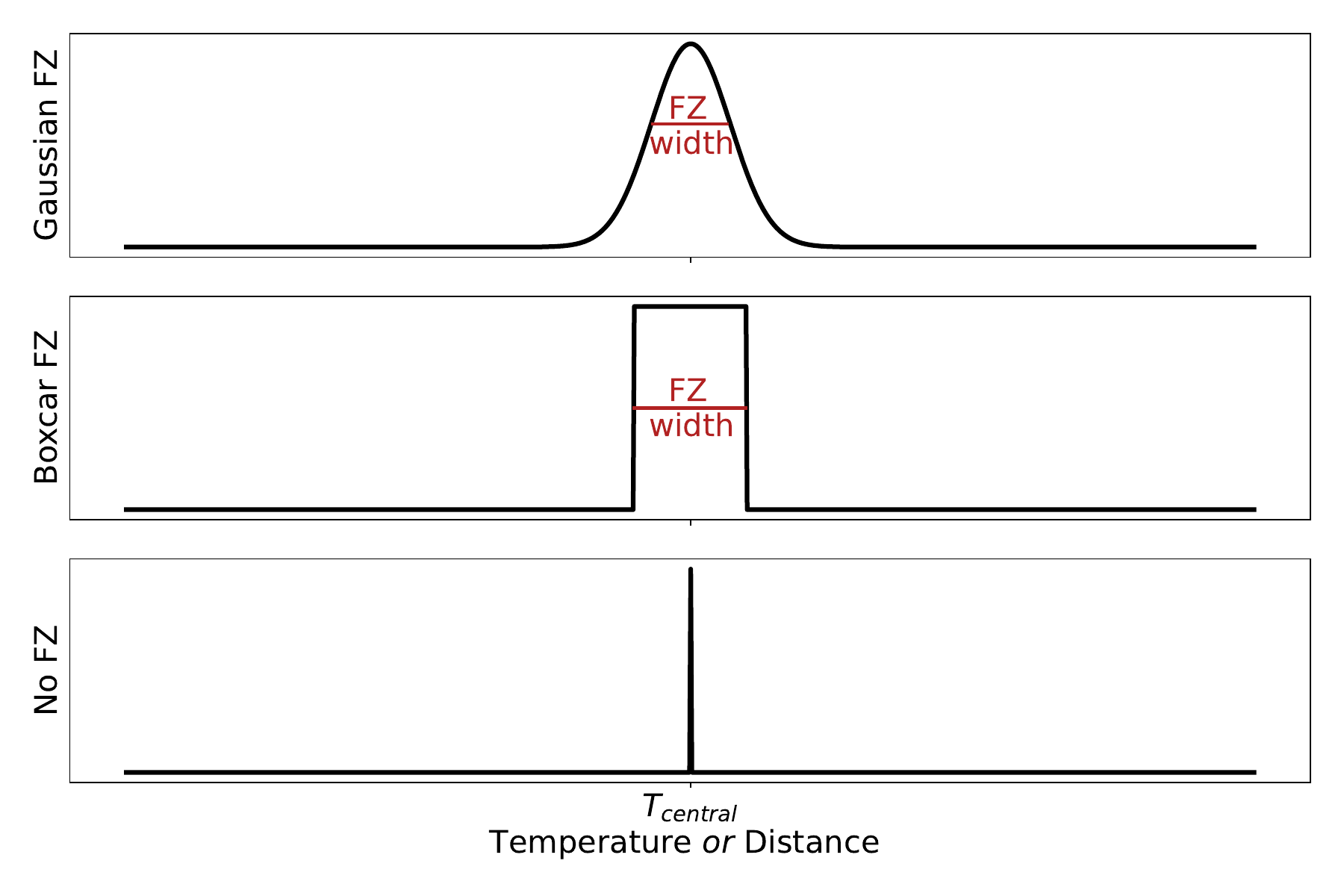}
    \caption{Illustration of the three different FZ models for creating a planet's composition at a given temperature. The x-axis denotes the temperature, or equivalently the distance from the star. \textit{Top panel}: Gaussian profile. \textit{Middle Panel}: Boxcar profile. \textit{Bottom Panel}: No FZ model.}
    \label{fig:fz_exp}
\end{figure}

Our underlying disk model is vastly simplified. The only parameter changing within our disk is the temperature, which is a proxy for distance from the central star. For our study, we are not interested in an exact temperature-distance relation but only in qualitative tendencies. We keep the pressure constant at a value of \SI{1e-4}{\bar}. This pressure value is often assumed for the formation of Earth \citep{Fegley2000,Lodders2003a,Wood2019}. In this context, however, the choice was arbitrary. As shown above in Sect. \ref{sec:condT_var_p}, variations in disk pressure affect the condensation of all species very similarly, especially if the expected variation in pressure is small.\footnote{Note that, in contrast to our simplified model, realistic disk models have a two-dimensional structure, show pressure gradients and pressure bumps; they likely have inhomogeneous element distributions, and they evolve in time.}

To derive the bulk composition of a planet for any given formation temperature (see above, Sect. \ref{sec:CompSol_Funcs_SimAnal}), we use three different methods of assembling planetary material to emulate planet formation via accretion. While our models are loosely rooted in the idea of planetary accretion from within the planet's Hill sphere \citep[see e.g.][]{Pollack1996,Kley1999}, we only use them to bypass computationally expensive $n$-body simulations. That means there is no correspondence between the expected physical size of the planet and the diversity of the material assembled in our code; we instead made the latter match the results of the $n$-body simulation. As illustrated in Fig. \ref{fig:fz_exp}, we compare two differently shaped planetary feeding zones (FZs) to a model without a FZ. 

In the simplest approach, the planet is only made up of the solid material that is stable at the planet's formation temperature, with the relative amounts dictated by the thermochemical equilibrium at that temperature. This approach corresponds to taking an infinitesimally thin section of the elements-temperature-progression described in Sect. \ref{sec:CompSol_Funcs_SimAnal}. It is illustrated in the bottom panel of Fig. \ref{fig:fz_exp}, where the $x$-axis represents the temperature decreasing with distance from the central star, all material except that at $T=T_{\rm{central}}$ is discarded. 

The first FZ is an equal weights temperature band, illustrated in the middle panel of Fig. \ref{fig:fz_exp}, later referred to as `boxcar FZ'. We specify the width of the temperature band, add up the elemental equilibrium compositions within the temperature range, and normalise the result. This normalised result is taken to be the planetary composition at the central temperature of the band. Since a lower temperature generally entails a higher total amount of solids in the equilibrium, it follows that the lower-temperature edge of the band effectively has a stronger influence on the resulting planetary composition than the higher-temperature edge. 

The second type of FZ is a Gaussian profile, as illustrated in the top panel of Fig. \ref{fig:fz_exp}. Here, the total material at each temperature is first multiplied by a normal distribution with a specified standard deviation, $\sigma$, and subsequently added up. The argument regarding more solid material being present at lower temperatures also applies to this FZ. The width of the FZ is given by $2\sigma$. The location of the peak of the normal distribution gives the planetary formation temperature. 

For both of these types of FZ, the effect of a ring geometry on the amount of available material is not taken into consideration for the final planetary make-up. That means we neglected the fact that a lower temperature corresponds to a greater distance from the star. A larger radius of the ring implies more material with that particular composition being available for accretion onto the planet. To quantify this geometric effect, we would have to connect our temperature profile to specific distances, which is not part of our simplified model.

\subsection{Application and comparison study}
Following the approach of B10, we analyse the predicted planetary compositions with our simplified disk model for chemically diverse stars, delineated primarily by differences in their \ce{C}/\ce{O} ratios. In particular, we explore the simulated compositions of a rocky planet formed around a low-carbon star (HD27442), around a medium-carbon star (HD17051), and around a high-carbon star (HD19994), using the three different planetary FZ models described above. The physical properties of the stars are listed in Table \ref{tab:Bond_stars}. We use the elemental abundances of these stars as reported by B10, in order to facilitate the comparison of the results. It should be noted, though, that these abundances have later been found to be inaccurate, especially the \ce{C}/\ce{O} ratios are vastly overestimated \citep{Fortney2012,Nissen2013,Teske2014,Brewer2016a}. Based on more recent studies of stars in the solar neighbourhood \citep[e.g.][]{Brewer2016}, all of the \ce{C}/\ce{O} ratios analysed in this section would be classified as moderately high or high. 

\begin{table}[]
\caption{Physical parameters of stars analysed in this section.}
\resizebox{0.5\textwidth}{!}{%
    \centering
    \begin{tabular}{llll}
    \hline\hline
         & HD 27442 & HD 17051 & HD 19994 \\\hline
        $T_{\rm{eff}}$ [\si{\K}] & 4825 & 6097 & 6188\\
        $M_{*}$ [$M_{\odot}$] & 1.48 & 1.15 & 1.37\\
        $R_{*}$ [$R_{\odot}$] & 3.43 & 1.18 & 1.75\\
        $L_{*}$ [$\log_{10}(L_{\odot})$] & 0.838 & 0.250& 0.626\\
        $[\ce{Fe}/\ce{H}]$ [dex] & 0.42   & 0.11  & 0.19 \\\hline
    \end{tabular}}
    \tablefoot{Values according to \citet{Turnbull2015}, as reproduced in the \href{https://exoplanetarchive.ipac.caltech.edu}{NASA Exoplanet Archive}.}
    \label{tab:Bond_stars}
\end{table}

We compare our results against the results of B10. They simulated the composition of rocky planets by combining a chemical equilibrium condensation with a dynamical accretion simulation. The chemical equilibrium condensation was done with the \textsc{HSC} suite. The $T$-$p$ input parameters were based on the \citet{Hersant2001} temperature-pressure-profile for the midplane of the protoplanetary disk.\footnote{\label{fn:diskprofile}The \citet{Hersant2001} disk profile is not considered state-of-the-art anymore, as it is a purely diffusive model. It has been shown that introducing radiative transfer to the model inverses the vertical temperature profile of the disk (i.e. $T$ increases for increasing $z$) compared to a purely diffusive model in which the temperature decreases with distance from the midplane, and a shadowing effect results in an overall cooler midplane \citep{Pinte2009,Woitke2009,Oberg2022}. However, for the purpose of our analysis, the only essential aspect of the disk-profile is that the midplane temperature decreases with distance from the star.} The solids formed in equilibrium at the specified temperature and pressure constitute the planetary building blocks for the dynamical $n$-body simulation, which was done using the \textsc{SyMBA} integrator \citep{Duncan1998}. They ran four accretion simulations for each of the studied stars, slightly varying the initial distribution of planetesimals, and recorded the composition of the final planets as the sum of all accreted material. Each simulation run returned between zero and three planets per star. For each star, we show the planet compositions in order of formation distance across the whole set of the B10 simulations. This illustrates representative compositional patterns as a function of distance, and captures their variability due to dynamics. 

\subsection{Results for chemically diverse systems}
Figures \ref{fig:HD27442_bc} to \ref{fig:HD19994_bc} show the comparisons among our results of planet compositions for the three stars and to the B10 results. We describe them in more detail in the following subsections. Each figure contains four panels, showing the bulk composition of a planet (in $\text{wt}-\%)$ simulated to form around the respective star. The top panel shows the discrete results of the B10 simulation as a function of distance, the second panel from the top shows our composition for a Gaussian FZ, the third panel shows our simulation for a boxcar FZ, and the bottom panel shows the composition without any FZ. Where possible, we group the B10 planets with similar composition and indicate roughly which section of our simulation best corresponds to them with arrows between the top most and second panel.

\subsubsection{Low carbon abundance}\label{sec:HD27442}
\begin{figure*}[ht!]
 \centering
 \includegraphics[width=0.7\textwidth]{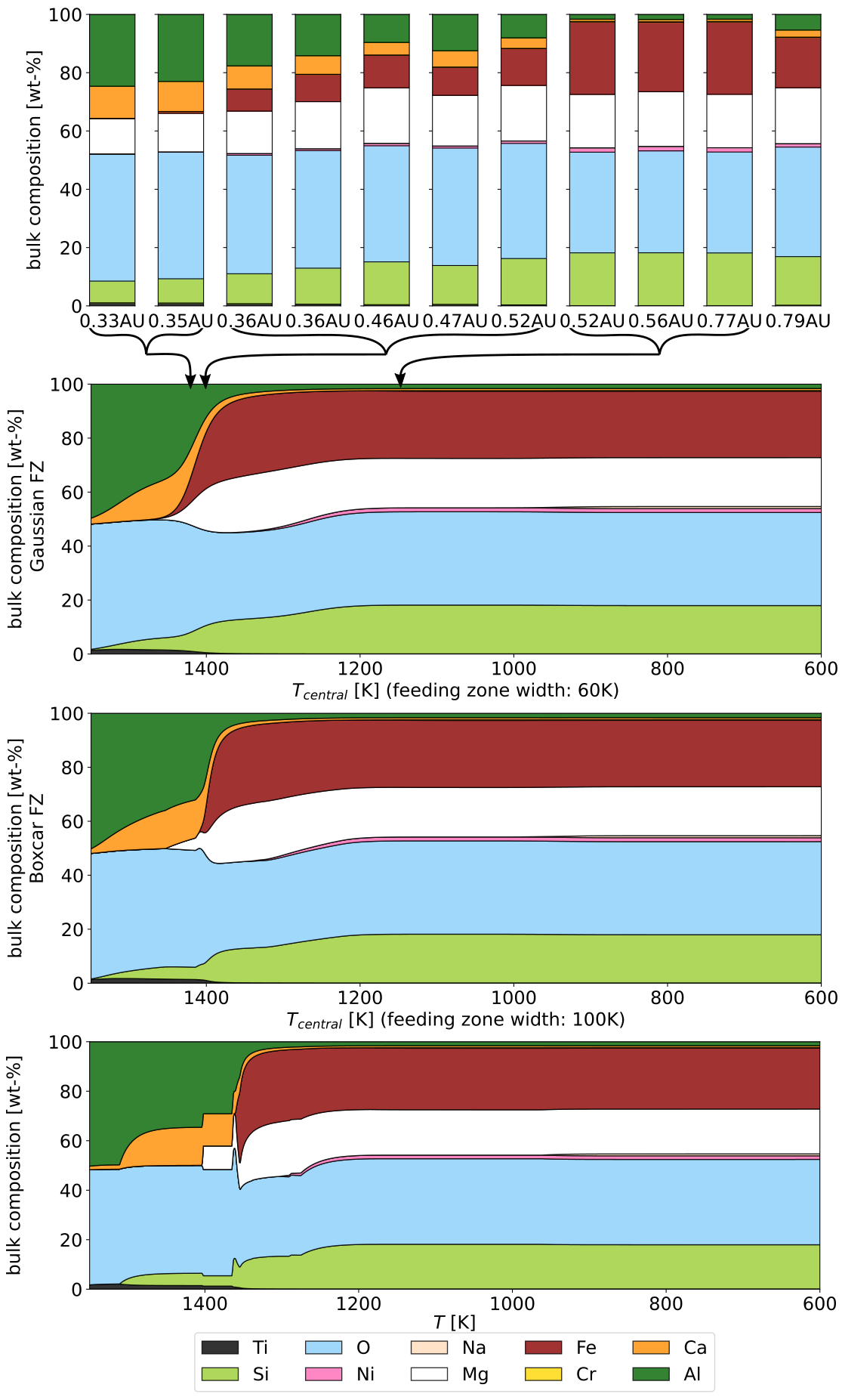}
 \caption{Predicted bulk composition (in $\text{wt}-\%$) of a rocky planet simulated for the elemental abundance of HD27442 (low carbon system). \textit{Top panel}: B10 planet composition results from four separate simulation runs. We also show our simulations: with a Gaussian FZ (\textit{second panel}), a boxcar FZ (\textit{third panel}), and no FZ (\textit{bottom panel}). The arrows between the first two panels indicate roughly the location of the best correspondence between the Bond simulation and ours.}
 \label{fig:HD27442_bc}
\end{figure*}

In Fig. \ref{fig:HD27442_bc} we show the system of HD27442, which has the lowest carbon abundance of the three analysed systems ($\ce{C}/\ce{O}=0.61$). The elemental abundance pattern of this star is similar to the solar values. This implies that the simulated planets can be expected to also resemble the inner planets of the Solar System in bulk composition. No \ce{S} abundance is reported for HD27442. Since its $\ce{C}/\ce{O}$ ratio is far below $1$, \ce{S} species are not expected to play a major role in the equilibrium chemistry (see footnote \ref{fn:S-O} and Sect. \ref{sec:HD19994}). We therefore excluded all species containing \ce{S} from the simulation of this system.

Starting closest to the central star, two planets from the B10 simulations formed at $0.33$ and \SI{0.35}{AU} with similar compositions of mostly \ce{O} and \ce{Al}, significant amounts of \ce{Mg}, \ce{Ca}, and \ce{Si}. We find a very similar composition in our simulation without a FZ at or slightly below \SI{1400}{\K}. The boxcar FZ, with a width of \SI{100}{\K}, produces a slightly different composition for this formation temperature, with a reduced \ce{Mg}-content. This is caused by mixing in material from the higher-temperature regions, where only \ce{Al}-\ce{Ca}-\ce{O} species have condensed. We cannot reproduce the composition of the two innermost planets with our Gaussian profile, because it does not create a region that contains \ce{Mg} but no \ce{Fe}. 

At intermediate distances, we find a group of five planets in the B10 simulation with similar \ce{O} and \ce{Si} contents as the innermost planets, but with ever increasing \ce{Fe} amounts. These planets formed between \SI{0.36}{AU} and \SI{0.52}{AU}, which seems to correspond to the location \ce{Fe} snow line in the B10 simulation. Due to the very abrupt condensation of \ce{Fe} at $T\approx \SI{1360}{\K}$, we cannot reproduce this planetary composition without resorting to a FZ. The gradual change in planet composition can be reproduced with both FZ models. The boxcar model would profit from a larger FZ width than the one used here, though.

At greater distances, we find a group of three planets in the B10 simulation that can be characterised by their large content of \ce{Fe}, \ce{Mg}, \ce{O}, and \ce{Si}. These planets formed between \SI{0.52}{AU} and \SI{0.77}{AU}. As shown in our continuous simulations, the composition of the solids in the disk converges to these specific ratios, which are in accordance with the stellar elemental abundance ratios. This is due to the fact that we now look at planetary formation temperatures below the condensation temperatures of the main planetary components. Once we enter this region, the FZ model becomes obsolete, as the material to either side of the central temperature is identical.

There is one final planet left in the B10 simulation that has no correspondence with our continuous simulation. It formed at the greatest distance from the star, but its composition rather resembles the second group of planets. The formation of this planet requires substantial dynamical processes, likely in the form of planet migration, which cannot be emulated by our simple FZ model.

\subsubsection{Medium carbon abundance}\label{sec:HD17051}
\begin{figure*}[ht!]
 \centering
 \includegraphics[width=0.7\textwidth]{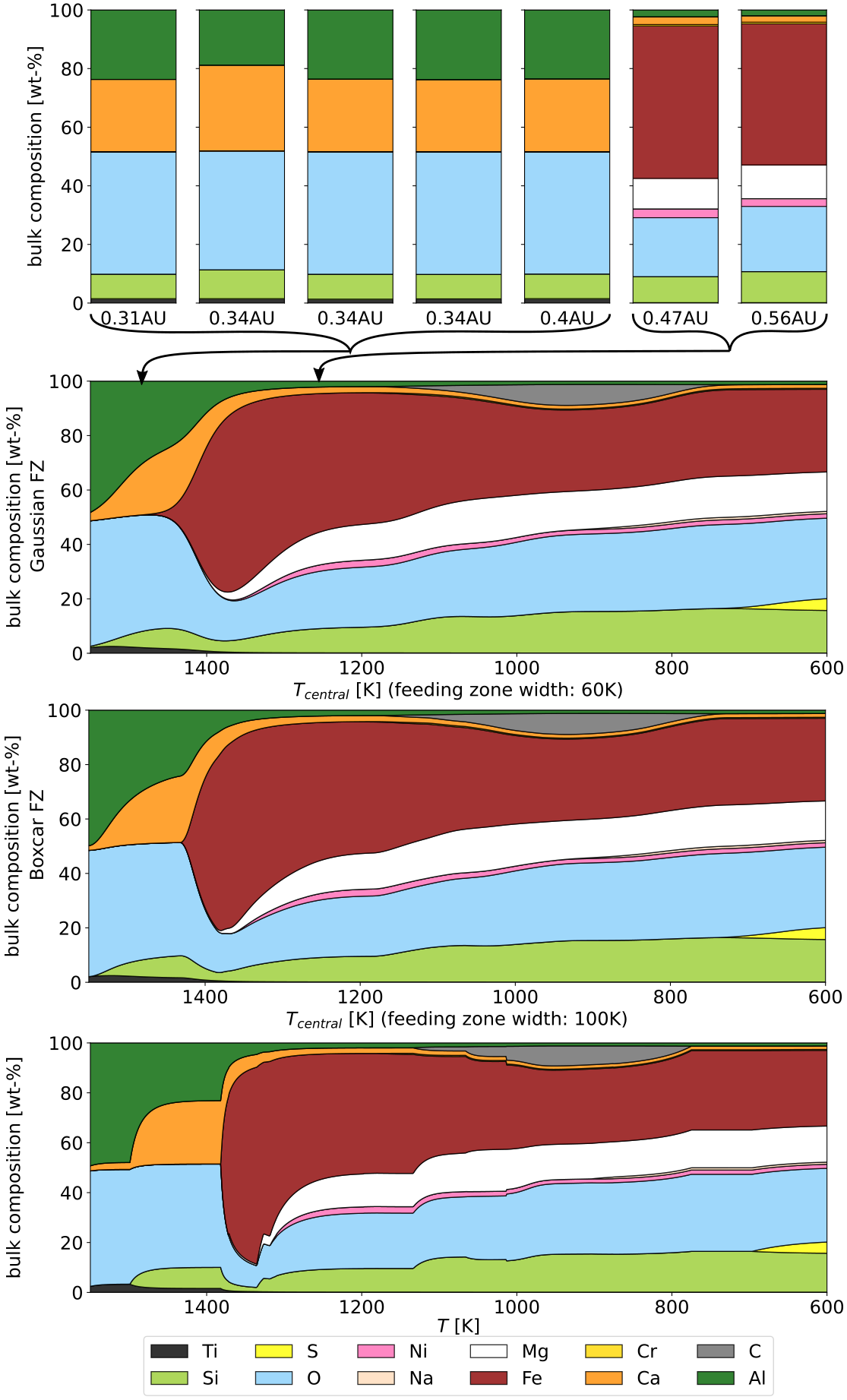}
 \caption{Predicted bulk composition (in wt-\%) of a rocky planet simulated for the elemental abundance of HD17051 (medium carbon system). \textit{Top panel}: B10 planet composition results from four separate simulation runs. We also show our simulation: with a Gaussian FZ (\textit{second panel}), with a boxcar FZ (\textit{third panel}), and with no FZ (\textit{bottom panel}). The arrows between the first two panels indicate roughly the location of the best correspondence between the Bond simulation and ours.}
 \label{fig:HD17051_bc}
\end{figure*}

In Fig. \ref{fig:HD17051_bc} we show the planets simulated to form around the star HD17051, which has an intermediate carbon abundance among the analysed systems ($\ce{C}/\ce{O}=0.87$). Closest to the central star, we see five B10 planets with almost identical compositions. They formed between approximately \SI{0.3}{AU} and \SI{0.4}{AU}. They contain a large fraction of \ce{O}, similar amounts of \ce{Al} and \ce{Ca}, and a small amount of \ce{Si}. We find the same composition in our simulation for all three types of FZs between roughly \SI{1500}{\K} and \SI{1400}{\K}. For the Gaussian FZ, though, the region corresponding to the composition of the B10 planets is very narrow, which is difficult to reconcile with the consistent compositions returned by the $n$-body simulation. The model without a FZ has a broad plateau of the same composition as the B10 planets. The boxcar FZ does not have such a broad plateau, but shows a section with a sufficiently constant composition to be compatible with the formation of similar planets over an extended range of distances. 

At greater distances, we identify two B10 planets with very similar composition that are vastly different from the first group. These planets are dominated by their high \ce{Fe} content, exceeding 50\% of the total weight of the planet, and suggesting a very extensive planetary core. The remaining composition is made up of \ce{O}, \ce{Mg}, and \ce{Si}, with only small contributions of \ce{Ni}, \ce{Ca}, and \ce{Al}. In contrast to the HD27442 system, this group of planets has not formed in region of the convergence composition of the disk. We can clearly see in our simulations that the composition changes significantly all the way down to approximately \SI{600}{\K}, when \ce{S} condenses. 

Both the high similarity of composition over a fairly large distance range, as well as the deviation from it in the form of a slight increase in \ce{Mg}, \ce{O}, and \ce{Si} can be seen in our three continuous models in the temperature range from approximately \SI{1300}{\K} to \SI{1200}{\K}. Both the Gaussian and the boxcar profile reproduce the gradual changes in the composition of the B10 planets. The composition without a FZ compares less favourably, because at the onset of the \ce{Mg} and \ce{Ni} condensation, it changes are very rapidly and strongly, and at lower temperatures, it stays constant.

\subsubsection{High carbon abundance}\label{sec:HD19994}
\begin{figure*}[ht!]
 \centering
 \includegraphics[width=0.7\textwidth]{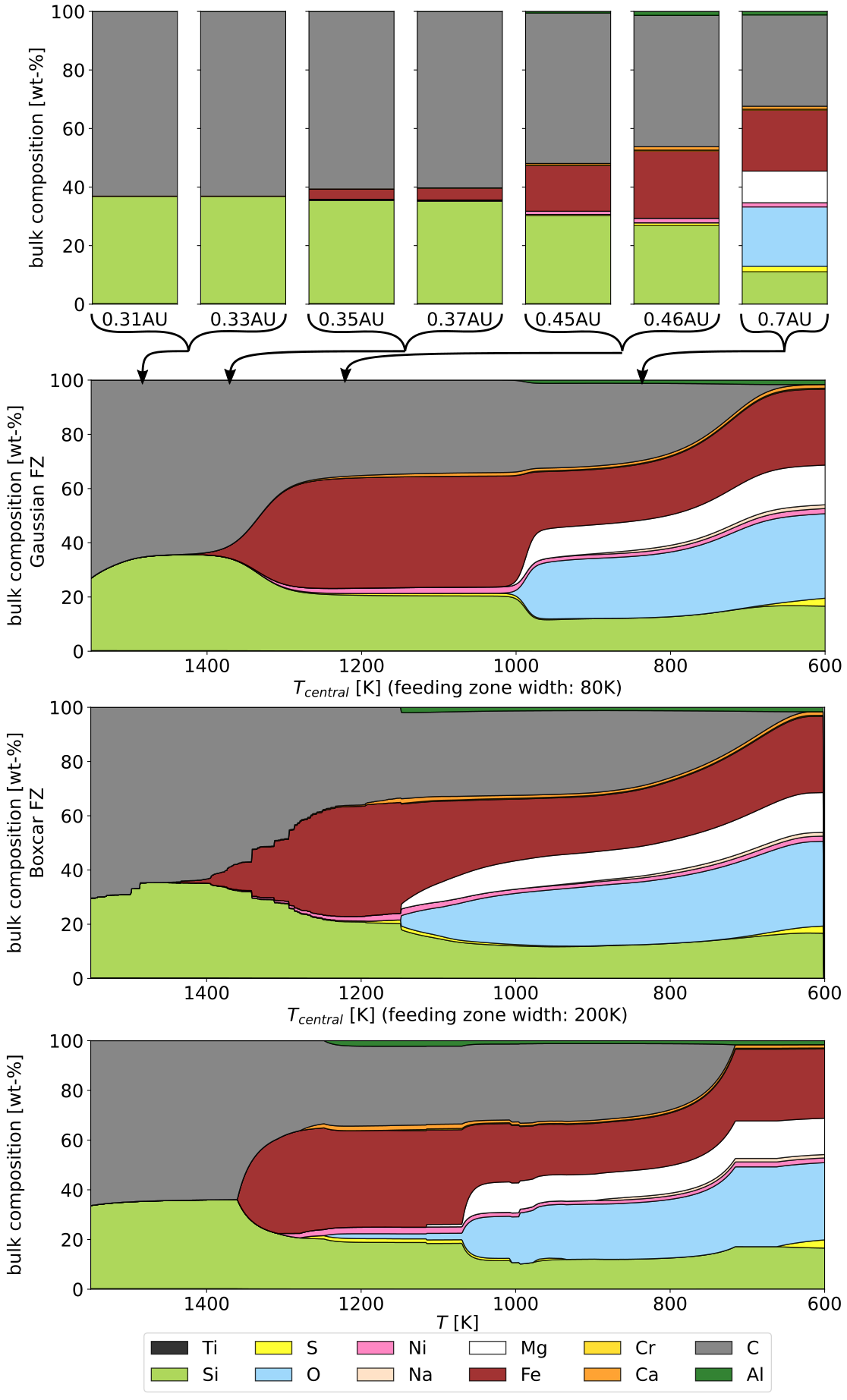}
 \caption{Predicted bulk composition (in wt-\%) of a rocky planet simulated for the elemental abundance of HD17051 (high carbon system). \textit{Top panel}: B10 planet composition results from four separate simulation runs. We also show our simulation with a Gaussian FZ (\textit{second panel}), with a boxcar FZ (\textit{third panel}), and with no FZ (\textit{bottom panel}). The arrows between the first two panels indicate roughly the location of the best correspondence between the Bond simulation and ours.}
 \label{fig:HD19994_bc}
\end{figure*}

Finally, we show the planets simulated to form around the high-carbon star HD199944 in Fig. \ref{fig:HD19994_bc}. Based on the elemental abundance data we use for this simulation, the star has the exceptional \ce{C}/\ce{O} ratio of $1.26$. We expect a completely altered disk chemistry for systems with \ce{C}/\ce{O} ratios exceeding unity. All \ce{O}-atoms are bound to \ce{C}-atoms to form highly stable \ce{CO} gas molecules \citep{Molliere2015,Woitke2018}. Accordingly, \ce{O} is no longer available for the solid-phase chemistry, inhibiting the condensation of some of the most common species in planet formation, such as \ce{Al2O3}, \ce{CaAl12O19}, and \ce{MgSiO3}. Because all \ce{O} is bound to \ce{CO}, no \ce{O} is available to bind with \ce{H2} to form \ce{H2O} and the system becomes highly reducing. This means that all \ce{Fe} is in reduced state and that some \ce{Si} occurs in metal instead of silicates. A \ce{C}/\ce{O} ratio exceeding $1$ also means that free \ce{C} is available to form exotic phases like \ce{SiC} or free \ce{C} in form of, for example, graphite. In high \ce{C}/\ce{O} systems, \ce{S} replaces \ce{O} as anion, which leads to a much higher condensation temperature of S as it condenses with \ce{Ca} into refractory phases like oldhamite (\ce{CaS}). Although the solar \ce{C}/\ce{O} is approximately 0.5, some portions of the early Solar System apparently had \ce{C}/\ce{O} ratios close to $1$ as it is evident from the presence of \ce{CaS} in reduced enstatite chondrites or exotic elemental ratio patterns of the rare Earth elements in ordinary chondrite chondrules \citep{Pack2004}. A planet with a bulk $\ce{C}/\ce{O} > 1$ would certainly not allow the presence of liquid water and thus would likely be hostile for life.

From a practical, computational point of view, it should be noted that not many \ce{S} species are taken into account in condensation simulations, due to their limited importance in solar-like systems. For instance, B10 only considers the solid \ce{S} species \ce{FeS}, \ce{MgS}, and \ce{CaS}; we only added \ce{Al2S3} to this selection.\footnote{The \textsc{GGchem} code \citep{Woitke2018}, on the other hand, contains thermochemical data of $12$ different solid \ce{S} species.} This means that the simulations likely do not reflect the true disk chemistry of a \ce{C}-rich system.

As expected, the composition of the simulated planets is completely different from the planets discussed so far. Starting again closest to the central star, we find two B10 planets only containing \ce{C} and \ce{Si}. They formed at $0.31$ and \SI{0.33}{AU}. In all our models, the same composition can be found for a large range of temperatures. The condensation of \ce{C} and \ce{SiC} in this system occurs at a much higher temperature than any of the other species, and in combination with the suppression of the \ce{Al}-\ce{Ca}-\ce{O} species, this \ce{C}-\ce{Si} composition of solids is very stable in the disk for an extended temperature range. This implies that the FZ type has hardly any influence on the predicted composition of the innermost planets in the system.

At intermediate distances, we find two B10 planets with a small fraction of \ce{Fe}. They formed at $0.35$ and \SI{0.37}{AU}. We cannot reproduce this composition without using a FZ. The very rapid condensation of \ce{Fe} means that the disk composition changes from no \ce{Fe} in solid form to all \ce{Fe} in solid form within a few kelvins. This makes it difficult to form a planet with a small amount of \ce{Fe}.

The next two B10 planets in the sequence, formed at $0.45$ and \SI{0.46}{AU}, show increasing amounts of \ce{Fe} and traces of other elements. Despite their almost identical distance from the central star, the ratios of these elements are substantially different. While we can identify a section in our FZ models, in which the \ce{Fe} fraction increases rapidly, we do not find these exact compositions in any of our models. Especially the \ce{Al} traces cannot be reproduced, as we found \ce{Al} to condense at a temperature that is too low to allow for mixing of the material into the region in which \ce{Fe} has not yet fully condensed. One reason for this deviation might be the geometric effect we described in Sect. \ref{sec:planet_comp_theory}. This would increase the relative amount of the lower-temperature material, making it available for a redistribution to the higher-temperature regions. It is also possible that this composition requires a more dynamical accretion of different types of materials than we can emulate with our FZ models. 

The most distant B10 planet, at \SI{0.7}{AU}, has a much more diverse composition. This planet formed at a temperature at which \ce{CO} starts to lose its role as the dominant \ce{C} gas phase and is replaced by \ce{CH4}, removing \ce{C} from the solid phase and freeing \ce{O} for the condensation of more common rocky species. Accordingly, the relative \ce{C} and \ce{Si} contents are significantly reduced, but there is also a large fraction of the typical rock components of \ce{O} and \ce{Mg}. Additionally, \ce{S} becomes a more abundant trace element. Qualitatively, we find this composition in all our models, the only difference seems to be that our models predict a much lower relative \ce{S} abundance at the location at which there is still a significant amount of \ce{C} in solids.

\subsection{Implication for simplified planet formation models}
We learned several things in the comparison between combined thermochemical-dynamic model of B10 and our simplified continuous planet composition models, where the only free parameter was the disk temperature.

Firstly, since the analysed B10 planets were confined to a radial distance between approximately \SI{0.3}{AU} and \SI{0.8}{AU} from their central star, the variations in disk pressure in these simulations is only about one order of magnitude. As we show in Sect. \ref{sec:condT_var_p}, the condensation temperatures of the elements do not change significantly within one order of magnitude in pressure. This makes it unsurprising that we can recreate the B10 results so easily without a variable pressure input. 

Regarding the emulation of dynamical planet formation, a FZ is generally able to reproduce the results of an $n$-body simulation. The continuum compositions of all three systems show that we can distinguish sections in which the element ratios are fairly constant over a large temperature range, and section with rapid changes. The greatest variability in composition occurs in the vicinity of the condensation temperatures of the major planet-building elements \ce{Mg}, \ce{Si}, and \ce{Fe}. At these temperatures, using a FZ is crucial to reproduce the gradual variations in planet composition found in $n$-body simulations. In regions where the element ratios are constant over a large temperature range, using a FZ is less relevant, or, in the case of the convergence composition, completely obsolete.

The exact shape of the FZ does not seem to be particularly significant, as their width can be adapted to generate the required effect on the final composition. For instance, \citet{Sossi2022} has shown that the measured elemental depletion pattern of Earth compared to the Sun can be achieved by using a Gaussian FZ with a standard deviation of approximately $\sigma \approx \SI{216}{\K}$, whereas that of Vesta, with its mass of $\SI{4e-5}{M_{\oplus}}$, requires a standard deviation of $\sigma \approx \SI{57}{\K}$. There are, however, some arguments in favour of the boxcar model. On the one hand, it seems to be better at reproducing the composition of the innermost planets formed in the $n$-body simulation. At the onset of condensation, when there is no solid material at higher temperatures, a Gaussian profile results in a very asymmetric assemblage of material that is skewed towards low-temperature material. On the other hand, a boxcar profile seems to be more compatible with the physical concept of accretion from a region within the gravitational influence of the forming planet. This could also be achieved by cutting off the wings of the Gaussian profile, for example at $2\sigma$ or $3\sigma$.

We have, however, seen some deviations from our continuum composition in the B10 planets, which we could not reproduce with any of our FZ models, and which must therefore be a result of the dynamical accretion simulation. This shows the limitation of our simplistic model. $N$-body simulations can help us explore the extent to which processes that entail large displacements of planetary building blocks from their formation region might affect the final composition of a planet. Taking this idea even further, these simulations would also allow us to study the composition of planets that are partly formed by accreting material from remote reservoirs \citep[`pebble accretion'; see e.g.][]{Kleine2020,Schneeberger2023,Gu2023}.

\section{Summary and conclusions}\label{sec:summary}
\textsc{ECCOplanets} is a simple, accessible, and versatile Python code that can be used to simulate the equilibrium condensation of the main building blocks of rocky planets in the protoplanetary disk of stars, as a function of the elemental abundance pattern and disk pressure, based on a Gibbs free energy minimisation. The performance of our code is stable and robust for a variety of starting conditions. The software package, which we make publicly available, includes a limited built-in (and extendable) library of thermochemical data representative of common problems in exoplanet formation.

In this paper we have used our code for two typical applications in planetary science: finding the condensation temperature of elements and condensates, and deriving the composition of rocky planets as a function of the stellar abundance pattern. Both these analyses were also used as a benchmark test for the results of our code against literature values. 

The computed condensation temperature of a condensate is very sensitive to its exact definition and to the selection of molecules included in the simulation. In combination with the uncertainty in thermochemical data, this suggests that the exact value of simulated molecular condensation temperatures is not very meaningful. Nevertheless, under reasonably simple assumptions, we have shown that our code outputs condensation temperatures within \SI{50}{\K} of accepted literature values for most tested species. 

The derived 50\% condensation temperatures of elements are a far more robust measure of disk chemistry. They are unambiguously defined and less sensitive to the selection of molecules. Here, the agreement between our results and the literature values is of the order of \SI{5}{\K}. The condensation temperatures of elements are highly sensitive to physical variations in the system, that is, the disk pressure and elemental abundance pattern.

The disk pressure affects the condensation temperature of all elements in a similar way, with higher pressures corresponding to higher condensation temperatures. Over the analysed range $10^{-6}$ to $10^{-1}$ \si{\bar}, we find an average increase in condensation temperatures of \SI{357\pm57}{\K} for the studied elements. 

To understand the influence of variations in the elemental abundance pattern, we performed simulations with synthetically altered key element ratios and compared them to a representative selection of stars. We identified different groups of systematic variations to the condensation temperatures, which hint at different underlying chemical processes. Regarding the number of affected elements and the magnitude of the change in condensation temperature, the metallicity and \ce{C}/\ce{O} ratio have the greatest impact. An increase in metallicity results in a log-linear increase in elemental condensation temperatures; in contrast, an increase in \ce{C}/\ce{O} lowers the condensation temperature exponentially. While not all elements are affected to the same degree, the condensation temperatures can easily vary by more than \SI{100}{\K} for the sampled parameter ranges of $4\times10^{-4}$ to $2\times10^{-3}$ in metallicity and $0.1$ to $0.7$ in \ce{C}/\ce{O}.

We conclude that the combined effect of the pressure and elemental abundance pattern on the condensation temperature of elements limits the applicability of the values derived in the context of the formation of the Earth to other planet formation locations within the Solar System, and especially other stellar systems. 
Finally, we studied the composition of rocky planets forming around three exemplary stars, delineated by their \ce{C}/\ce{O} ratio. To explore the effects of profoundly limited model assumptions, we used a one-parameter ($T$) disk model and only emulated planetary accretion with FZ models. We compared our results against a study using a ($T$-$p$) disk model in a combined thermochemical and $n$-body simulation. 

Our simple model was able to reproduce almost all compositions of the combined thermochemical-dynamical simulation. This serves as a further confirmation that the disk pressure has an almost uniform influence on the whole condensation regime, and that neglecting it does not affect the results qualitatively for small pressure ranges. It also provides insights into the effects of dynamical accretion. Dynamical accretion leads to gradual changes in the planetary composition as a function of distance from the star. As most elements condense abruptly, these gradual changes require the mixing of condensates from the equilibrium conditions of a large temperature range, that is, a FZ. The shape of the FZ appears to be insignificant, as any FZ can be tailored to achieve the required degree of redistribution of material by adjusting its width.

We conclude that the most likely main characteristics of rocky planet compositions can be determined with very simplified model assumptions. Adding further model parameters can give us invaluable insights into the variability and deviations from equilibrium conditions to be expected in a real exoplanet population.

% \bibliographystyle{aa} % style aa.bst
% \bibliography{Yourfile} % your references Yourfile.bib
\bibliographystyle{aa}
\bibliography{references}

\clearpage
\onecolumn

%\newpage
\begin{appendix}

\section{Example of  a stoichiometry matrix}\label{app:stoichiometry}
We consider a system only containing the elements \ce{H}, \ce{O}, \ce{C}. The initial amount of each element $i$ is denoted as $b_i$. 
We assume that these elements can only form the molecules \ce{H2}, \ce{O2}, \ce{H2O}, \ce{C}, \ce{CO2}, and \ce{CH4}. The amount of each of the molecules $i$ is denoted as $x_i$. \
Then, the number balance of the system requires\begin{align*}
b_{\ce{O}} &= 2 x_{\ce{O2}} + x_{\ce{H2O}} + 2 x_{\ce{CO2}}\\
b_{\ce{H}} &= 2 x_{\ce{H2}} + 2 x_{\ce{H2O}} + 4 x_{\ce{CH4}}\\ 
b_{\ce{C}} &= x_{\ce{C}} + x_{\ce{CO2}} + x_{\ce{CH4.}}\\ 
\end{align*}
\noindent
Alternatively, this equation can be written in vector notation as\begin{align*}
 \begin{pmatrix}
 b_{\ce{O}}\\
 b_{\ce{H}}\\
 b_{\ce{C}}
 \end{pmatrix}
 &=
 \begin{pmatrix}
 2 & 1 & 2 & 0 & 0 & 0 \\
 0 & 2 & 0 & 2 & 4 & 0\\
 0 & 0 & 1 & 0 & 1 & 1 
 \end{pmatrix}
 \cdot
 \begin{pmatrix}
 x_{\ce{O2}}\\
 x_{\ce{H2O}}\\
 x_{\ce{CO2}}\\
 x_{\ce{H2}}\\
 x_{\ce{CH4}}\\
 x_{\ce{C}}\\
 \end{pmatrix}\\
 \mathbf{b} &= \mathbf{A} \cdot \mathbf{x}.
\end{align*}

\newpage
\section{Simulation parameters} 

\begin{table}[ht]
\caption{Simulation parameters for performance and stability tests.}
 \centering
 \begin{tabular}{p{1.5cm}p{6.5cm}}
 \hline\hline
  $T_{start}$ & \SI{6000}{\K}, \SI{4000}{\K}, \SI{2000}{\K}, \SI{1700}{\K} \\
  $T_{end}$ & \SI{300}{\K} \\
  $\Delta T$ & \SI{1}{\K}, \SI{0.1}{\K}, \SI{5}{\K}, \SI{10}{\K} \\\hline
  system pressure & \SI{1e-4}{\bar}\\\hline
  elemental abundance & Solar \\\hline
  gas-phase species & \ce{Al},
\ce{Ti},
\ce{SiO},
\ce{Si},
\ce{O2},
\ce{MgO},
\ce{Mg},
\ce{H2O},
\ce{H2},
\ce{CO2},
\ce{CO},
\ce{CH4},
\ce{Fe},
\ce{CaO},
\ce{Al2O},
\ce{C},
\ce{Ca},
\ce{TiO}\\\hline

 solid-phase species & \ce{Fe3O4},
\ce{CaAl2Si2O8},
\ce{CaAl12O19},
\ce{Ca2Al2SiO7},
\ce{CaTiO3},
\ce{MgAl2O4},
\ce{TiC},
\ce{MgSiO3},
\ce{C},
\ce{CaAl4O7},
\ce{SiC},
\ce{Al2O3},
\ce{CaMgSi2O6},
\ce{Fe},
\ce{Mg2SiO4}\\\hline
 \end{tabular}
 \label{tab:SimParams_perf+stab}
\end{table}

\begin{table}[ht]
 \caption{Simulation parameters for literature comparisons.}
 \centering
 \begin{tabular}{p{1.5cm}p{6.5cm}}
 \hline\hline
  $T_{start}$ & \SI{2000}{\K} \\
  $T_{end}$ & \SI{300}{\K} \\
  $\Delta T$ & \SI{1}{\K} \\\hline
  system pressure & \SI{1e-4}{\bar}\\\hline
  elemental abundance & Solar \\\hline
  gas-phase species & 
  \ce{H}, \ce{H2}, \ce{He}, \ce{C}, \ce{CO}, \ce{CH4}, \ce{N}, \ce{N2}, \ce{NH3}, \ce{H2O}, \ce{O}, \ce{Fe}, \ce{OH}, \ce{SiO}, \ce{MgO2H2}, \ce{FeO2H2}, \ce{AlO2H}, \ce{CaO2H2}, \ce{AlOH}, \ce{Al2O}, \ce{Na2O2H2}, \ce{Na}, \ce{NaH}, \ce{NaOH}, \ce{Mg}, \ce{MgH}, \ce{MgOH}, \ce{Al}, \ce{AlH}, \ce{AlO}, \ce{Si}, \ce{SiH4}, \ce{SiS}, \ce{SiH}, \ce{S}, \ce{HS}, \ce{H2S}, \ce{NiS}, \ce{Ca}, \ce{CaOH}, \ce{Ti}, \ce{TiO}, \ce{TiO2}, \ce{Cr}, \ce{Fe}, \ce{FeS}, \ce{Ni}\\\hline

 solid-phase species & 
\ce{Al2O3}, \ce{Ca2Al2SiO7}, \ce{Mg2SiO4}, \ce{MgSiO3}, \ce{MgAl2O4}, \ce{Ni}, \ce{CaAl2Si2O8}, \ce{CaMgSi2O6}, \ce{NaAlSi3O8}, \ce{Ca2MgSi2O7}, \ce{MgO2H2}, \ce{Cr}, \ce{MgCr2O4}, \ce{CaTiO3}, \ce{Ti4O7}, \ce{CaTiSiO5}, \ce{FeS}, \ce{MgS}, \ce{CaS}, \ce{Al2S3}, \ce{FeTiO3}, \ce{C}, \ce{Fe}, \ce{SiC}, \ce{TiC}\\\hline
 \end{tabular}

 \label{tab:SimParams}
\end{table}

\newpage
\section{Stability and performance tests}\label{app:StabPerf}
We performed several simulations to assess the robustness and internal consistency of our results. In particular, we investigated in the influence of (1) the starting temperature, $T_{\rm{start}}$, (2) the temperature resolution, $\Delta T$, and
(3) he scaling of the elemental abundance pattern of the system.
 
The less noticeable the influence of these variations on the simulation result, the more reliable we judge them to be. In general, the quality of a simulation can be assessed by comparing the computed condensation temperatures against literature values, checking the smoothness of the condensation curves, that is, the lack of numerical errors.

We tested the starting temperatures $T_{\rm{start}} = \left[6000,\; 4000,\; 2000,\; 1700\right] \si{\K}$, that is, only temperature above the expected onset of condensation at $T \approx \SI{1670}{\K}$ for this simulation. The temperature resolutions covered the values $\Delta T = \left[0.1,\; 1,\; 5,\; 10\right] \si{\K}$. For the abundance scaling, we normalised all elements to the abundance of \ce{Si}, with $n_{\ce{Si}} = \left[10^5,\; 10^6,\; 10^7\right]$. If not otherwise specified, the simulations were run with a resolution of $\Delta T = \SI{1}{\K}$ and a start temperature of $T_{\rm{start}} = \SI{4000}{\K}$.

Our test problem includes 33 common species, divided into 15 solid-phase species and 18 gas-phase species, made out of the elements \ce{Fe}, \ce{O}, \ce{Al}, \ce{Ca}, \ce{Si}, \ce{Ti}, \ce{Mg}, \ce{C,} and \ce{H}. These elements, except for \ce{C} and \ce{H}, are the major constituents of the rocky planets in the Solar System. For the sake of simplicity, \ce{Ni}, which behaves similar as metallic \ce{Fe} and \ce{S}, which mainly occurs in the outer core, have been neglected here. We use both the Solar System relative elemental ratios and the presumed disk pressure at \SI{1}{AU} of \SI{1e-4}{\bar} reported by \citet{Lodders2003a}.\footnote{We ran many simulations with larger sets of species, containing more elements, without encountering any major computational problems, but did not perform any systematic performance tests.} The simulation parameters are summarised in Table \ref{tab:SimParams_perf+stab}.

We used two types of simulation results to quantitatively assess the robustness of the simulation: (1) the 50\% condensation temperature of elements and (2) the elemental composition of the solids in the system at three different temperatures ($T = \left[1600,\; 1400,\; 1200\right] \si{\K}$).

\begin{table}[ht]
\caption{50\% condensation temperature of different elements for variations in the T parameter.}
 \centering
 \resizebox{0.5\textwidth}{!}{%
 \begin{tabular}{lllllll}
 \hline\hline
  Simulation & \multicolumn{6}{c}{50\% condensation Temperature in \si{\K}} \\
   & \ce{Al} & \ce{Ca} & \ce{Mg} & \ce{Fe} & \ce{Si} & \ce{Ti}\\ \hline
  $T_{start} = \SI{6000}{\K}$ & 1653 & 1534 & 1336 & 1335 & 1317 & 1591 \\
  $T_{start} = \SI{4000}{\K}$ & 1653 & 1534 & 1336 & 1335 & 1317 & 1591 \\
  $T_{start} = \SI{2000}{\K}$ & 1653 & 1534 & 1336 & 1335 & 1317 & 1591 \\
  $T_{start} = \SI{1700}{\K}$ & 1653 & 1534 & 1336 & 1335 & 1317 & 1591 \\\hline 
  $\Delta T = \SI{0.1}{\K}$ & 1652.5 & 1534.3 & 1335.7 & 1335.1 & 1317.5 & 1590.5 \\ 
  $\Delta T = \SI{5}{\K}$ & 1655 & 1535 & 1335 & 1335 & 1315 & 1590 \\
  $\Delta T = \SI{10}{\K}$ & 1650 & 1530 & 1340 & 1330 & 1320 & 1590 \\\hline
 \end{tabular}}
 \label{tab:test_T_var_condT}
\end{table}

\begin{table}[ht]
 \caption{Elemental composition of solids at different temperatures as a function of variations in the T parameter.}
 \centering
 \resizebox{0.5\textwidth}{!}{%
 \begin{tabular}{llllllll}
 \hline\hline
  Simulation & \multicolumn{7}{c}{Elemental composition of solids in wt-\%} \\
   & \ce{Al} & \ce{Ca} & \ce{O} & \ce{Mg} & \ce{Fe} & \ce{Si} & \ce{Ti}\\ \hline
  \multicolumn{8}{c}{$T=\SI{1600}{\K}$} \\ \hline
  $T_{start} = \SI{6000}{\K}$ & 37.32 & 3.21 & 59.38 & 0 & 0 & 0 & 0.10 \\
  $T_{start} = \SI{4000}{\K}$ & 37.32 & 3.21 & 59.38 & 0 & 0 & 0 & 0.10 \\
  $T_{start} = \SI{2000}{\K}$ & 37.32 & 3.21 & 59.38 & 0 & 0 & 0 & 0.10 \\
  $T_{start} = \SI{1700}{\K}$ & 37.32 & 3.21 & 59.38 & 0 & 0 & 0 & 0.10 \\
  $\Delta T = \SI{0.1}{\K}$ & 37.32 & 3.21 & 59.38 & 0 & 0 & 0 & 0.10 \\
  $\Delta T = \SI{5}{\K}$ & 37.32 & 3.21 & 59.38 & 0 & 0 & 0 & 0.10 \\
  $\Delta T = \SI{10}{\K}$ & 37.32 & 3.21 & 59.38 & 0 & 0 & 0 & 0.10 \\\hline 
  \multicolumn{8}{c}{$T=\SI{1400}{\K}$} \\ \hline
  $T_{start} = \SI{6000}{\K}$ & 18.38 & 13.74 & 58.16 & 2.59 & 0 & 6.60 & 0.53 \\
  $T_{start} = \SI{4000}{\K}$ & 18.38 & 13.74 & 58.16 & 2.59 & 0 & 6.60 & 0.53 \\ 
  $T_{start} = \SI{2000}{\K}$ & 18.38 & 13.74 & 58.16 & 2.59 & 0 & 6.60 & 0.53 \\ 
  $T_{start} = \SI{1700}{\K}$ & 18.38 & 13.74 & 58.16 & 2.59 & 0 & 6.60 & 0.53 \\ 
  $\Delta T = \SI{0.1}{\K}$ & 18.38 & 13.74 & 58.16 & 2.59 & 0 & 6.60 & 0.53 \\
  $\Delta T = \SI{5}{\K}$ & 18.38 & 13.74 & 58.16 & 2.59 & 0 & 6.60 & 0.53 \\
  $\Delta T = \SI{10}{\K}$ & 18.38 & 13.74 & 58.16 & 2.59 & 0 & 6.60 & 0.53 \\\hline 
  \multicolumn{8}{c}{$T=\SI{1200}{\K}$} \\ \hline
  $T_{start} = \SI{6000}{\K}$ & 1.36 & 1.02 & 51.65 & 16.47 & 13.44 & 16.02 & 0.04 \\
  $T_{start} = \SI{4000}{\K}$ & 1.36 & 1.02 & 51.65 & 16.47 & 13.44 & 16.02 & 0.04 \\
  $T_{start} = \SI{2000}{\K}$ & 1.36 & 1.02 & 51.65 & 16.47 & 13.44 & 16.02 & 0.04 \\
  $T_{start} = \SI{1700}{\K}$ & 1.36 & 1.02 & 51.65 & 16.47 & 13.44 & 16.02 & 0.04 \\
  $\Delta T = \SI{0.1}{\K}$ & 1.36 & 1.02 & 51.65 & 16.47 & 13.44 & 16.02 & 0.04 \\
  $\Delta T = \SI{5}{\K}$ & 1.36 & 1.02 & 51.65 & 16.47 & 13.44 & 16.02 & 0.04 \\
  $\Delta T = \SI{10}{\K}$ & 1.36 & 1.02 & 51.65 & 16.47 & 13.44 & 16.02 & 0.04 \\\hline
 \end{tabular}}

 \label{tab:test_T_var_comp}
\end{table}

In Table \ref{tab:test_T_var_condT}, we summarise the result of our tests with regard to the condensation temperatures of elements. It is clear that neither the starting temperature, $T_{start}$, nor the temperature resolution, $\Delta T$, have an effect on the simulation. The only deviations we observe are due to the reduced/increased temperature sampling that is entailed by changing the temperature resolution of the simulation. In Table \ref{tab:test_T_var_comp} we summarise the second type of results: the relative elemental composition of solids at sample temperatures. The results were identical irrespective of the simulation parameters. 

We did observe some qualitative differences when examining the temperatures progressions of the test simulations, though. The chosen temperature resolution $\Delta T$ was most influential in this regard. The higher the resolution, the more stable the simulation's response to abrupt changes in gradients of the curves. Our simulations with low resolutions ($\Delta T > \SI{1}{\K}$) showed a tendency to overshoot significantly at these locations. Regarding the starting temperature, we found that irrespective of $T_{start}$, the simulations usually need a few temperature steps as a `burn in' period, in which the simulation results are unreliable. This suggests that a simulation should always be started at a temperature at least \SI{20}{\K} above the temperature range of interest. There were almost no differences in the simulation results as a function of the abundance scaling. Only for the normalisation $n_{\ce{Si}} = 10^7$ we found some isolated numerical errors (overshoots at sharp gradient changes), which we cannot explain.

\begin{figure*}[ht]
  \begin{subfigure}[t]{0.5\textwidth}
   \centering
   \includegraphics[width=\textwidth]{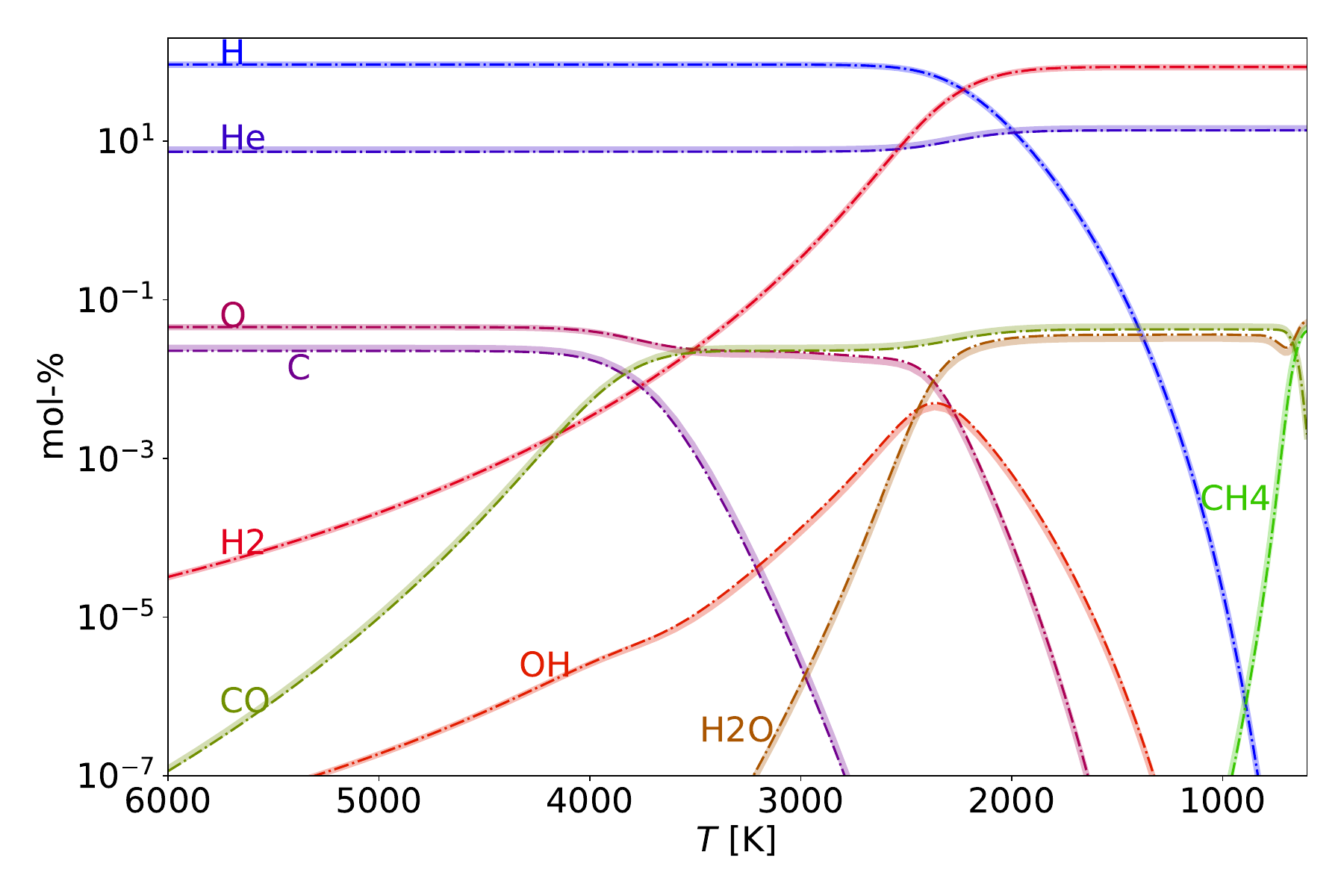}
  \end{subfigure}
  \hfill
  \begin{subfigure}[t]{0.5\textwidth}
   \centering
   \includegraphics[width=\textwidth]{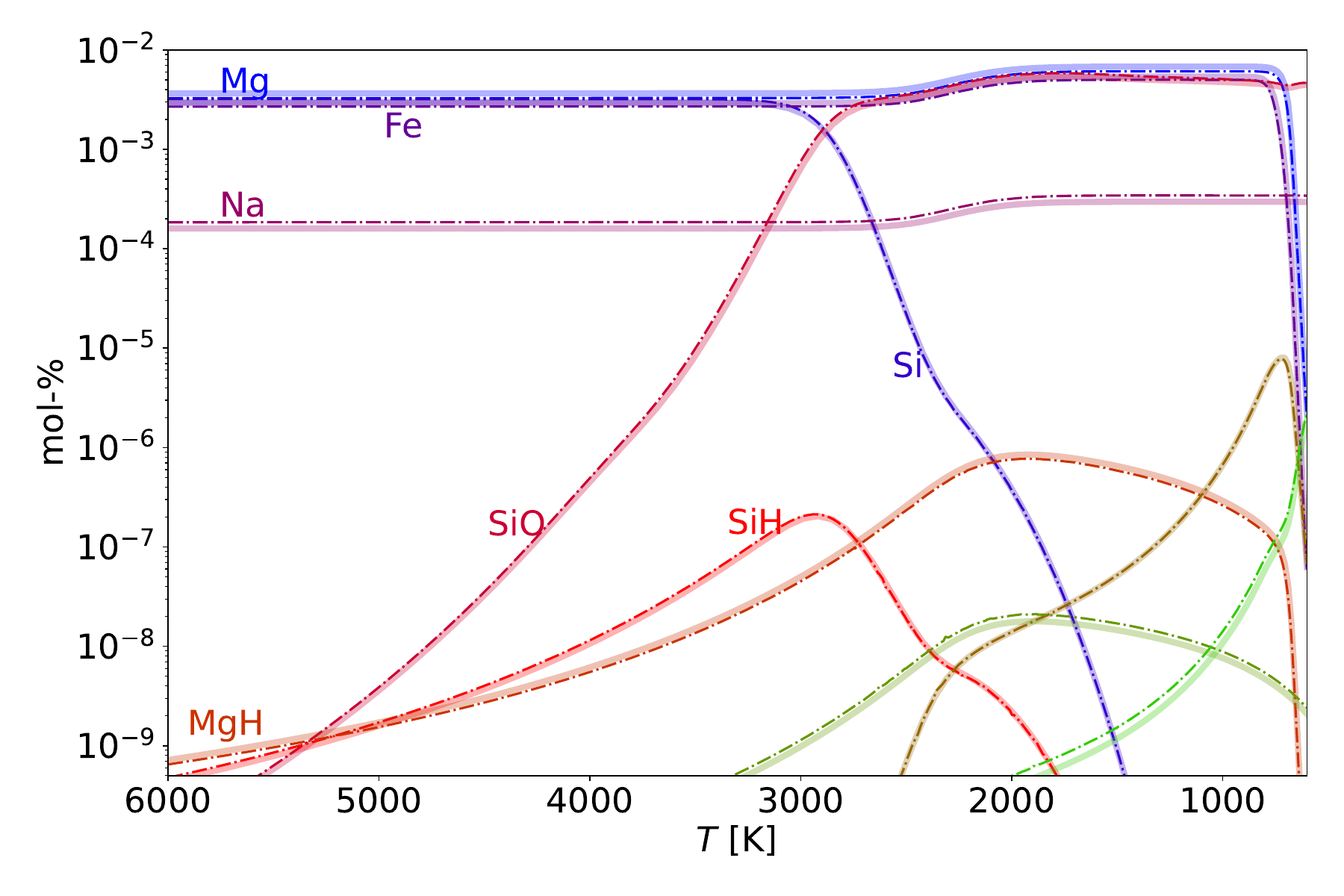}
  \end{subfigure}
  \begin{subfigure}[t]{0.5\textwidth}
   \centering
   \includegraphics[width=\textwidth]{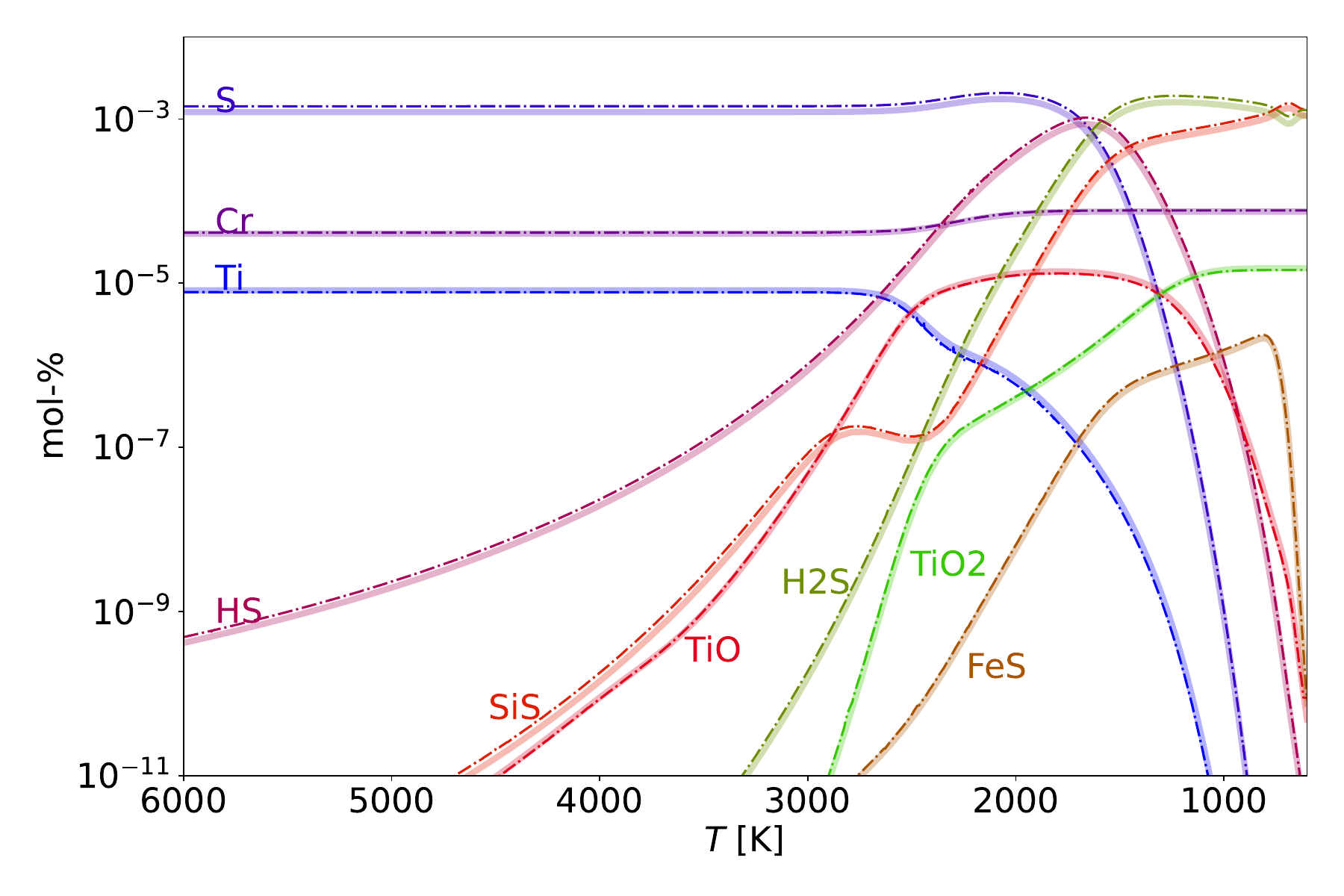}
  \end{subfigure}
  \hfill
  \begin{subfigure}[t]{0.5\textwidth}
   \centering
   \includegraphics[width=\textwidth]{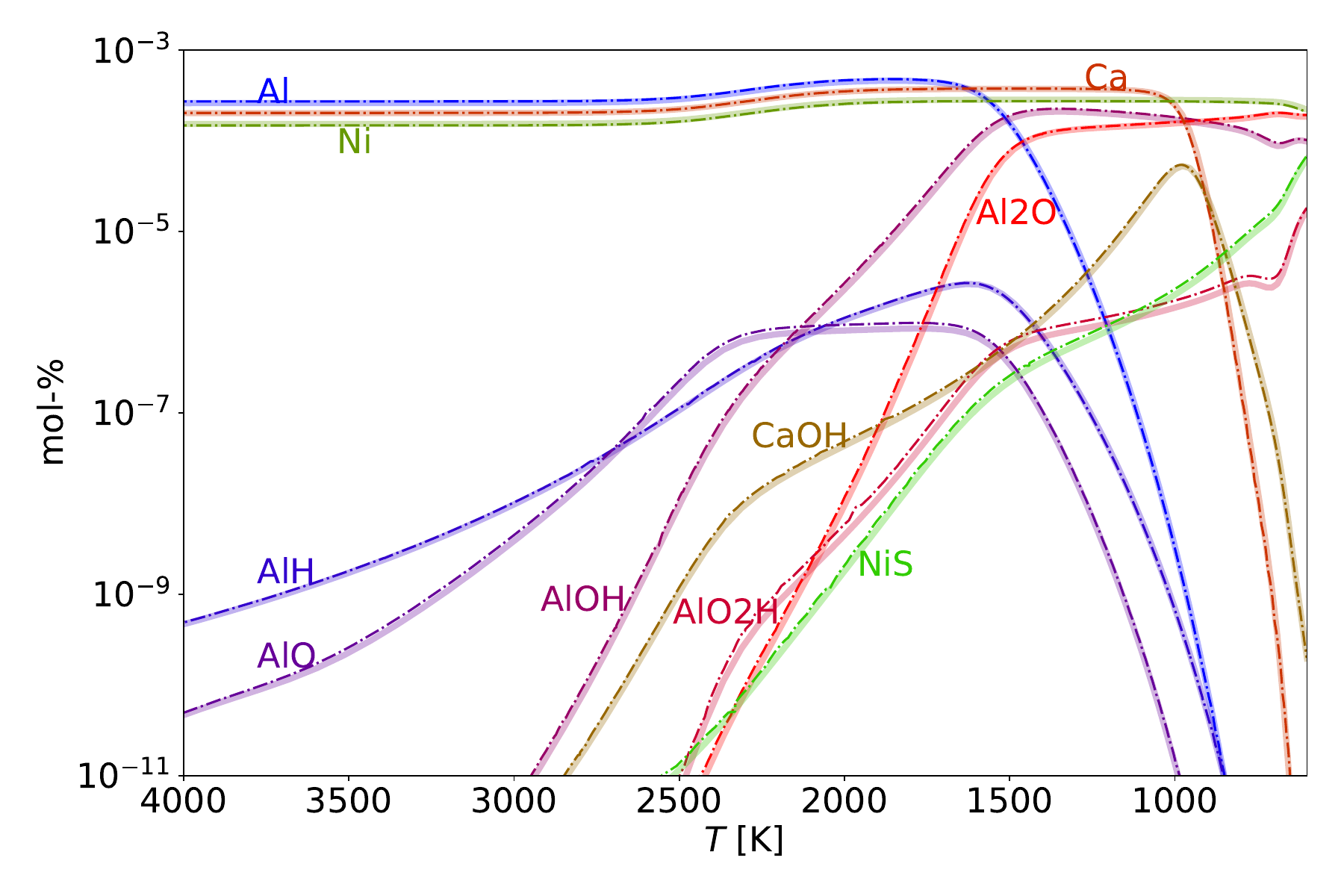}
  \end{subfigure}
  \caption{Benchmark test results of a gas-phase simulation. Light solid lines show \textsc{GGchem} output. Dash-dotted lines show our output. Both simulations were run at a constant pressure of $p=10^{-4}~\si{\bar}$, from $T=\SI{6000}{\K}$ to $T=\SI{600}{\K}$. The \textsc{GGchem} simulation included $120$ gas-phase species; our simulation included $40$. There were slight differences in the elemental abundance pattern assumed in the two simulations.}
  \label{fig:bench_GGchem}
\end{figure*}

Additionally, we perform a pure gas-phase benchmark test against the open-source condensation code \textsc{GGchem} by \citet{Woitke2018}, which itself is benchmarked against the open-source gas-phase code \textsc{Tea} by \citet{Blecic2016b}. In this test, we run both codes `as is', that is, we do not try to make the simulations as similar as possible, but use them in their default configuration, with regards to solar abundance pattern\footnote{\textsc{GGchem} uses the solar abundance pattern as reported by \citet{Asplund2009}.} and included number of species. We set the disk pressure to the same value in the two simulations and restrict both to the same set of elements.\footnote{The simulation details can be found in the git repository of \textsc{ECCOplanets}.}

The gas-phase benchmark test against \textsc{GGchem} shows an overall good agreement (see Fig \ref{fig:bench_GGchem}). The curve shapes are identical for all species included in both simulations, despite the large difference in the total number of included molecules. Most deviations in the relative amounts are explained by differences in the assumed elemental abundance pattern. We consider the found accuracy sufficient for all intended uses of our code.

\begin{table}[ht]
 \caption{Comparison of run times for different tests.}
 \centering
 \begin{tabular}{ll|p{2.5cm}p{2.5cm}}
 \hline\hline
  $T_{\rm{start}}$ & $\Delta T$ & total run time [h:mm:ss] & computation time per $T$-step [s] \\ \hline
  $\SI{6000}{\K}$ & \multirow{4}{*}{\SI{1}{\K}} & 0:48:03 & 0.506 \\
  $\SI{4000}{\K}$ & & 0:31:32 & 0.511 \\
  $\SI{2000}{\K}$ & & 0:14:52 & 0.525 \\
  $\SI{1700}{\K}$ & & 0:12:05 & 0.518 \\\hline 
  \multirow{4}{*}{\SI{4000}{\K}} & $\SI{0.1}{\K}$ & 5:25:12 & 0.527 \\
   & $\SI{1}{\K}$ & 0:31:32 & 0.511 \\
   & $\SI{5}{\K}$ & 0:06:40 & 0.541 \\
   & $\SI{10}{\K}$ & 0:04:32 & 0.735 \\\hline 
 \end{tabular}
 \tablefoot{Computer specifications: Intel(R) Core(TM) i7-8565U CPU @ 1.80GHz 1.99 GHz, RAM: 16.0 GB. Simulations run in Spyder 5.1.5 IDE.}
 \label{tab:runtime}
\end{table}

While the results of the stability tests are all in good agreement, the simulation parameters obviously affect the computation time of the simulation. As shown in Table \ref{tab:runtime}, the computation time per temperature step is very similar for all test except the lowest resolution of $\Delta T = \SI{10}{\K}$. Disregarding this simulation, we find a computation time per temperature step of $t=\SI{0.52 \pm 0.01}{\s}$. Accordingly, the total runtime scales linearly with the number of temperature steps to be calculated.

\newpage
\onecolumn
\section{Included molecule data}\label{app:molecules}

\begin{table}[ht]
 \caption{Included solid-phase data.}
 \centering
 \begin{tabular}{lll}
 \hline\hline
 chemical formula & phase & data source \\\hline
\ce{CaAl12O19(s)} & Hibonite & \citet{Allibert1981}\\
\ce{CaAl4O7(s)} & Grossite & \citet{Allibert1981}\\
\ce{Ca3P2O8(s)} & Tricalcium phosphate & \citet{Robie1995}\\
\ce{Ca2Al2SiO7(s)} & Gehlenite & \citet{Robie1995}\\
\ce{FeAl2O4(s)} & Hercynite & \citet{Robie1995}\\
\ce{CaAl2Si2O8(s)} & Anorthite & \citet{Hemingway1982}\\
\ce{SiO2(s)} & Quartz & NIST-JANAF\\
\ce{FeTiO3(s)} & Ilmenite & \citet{Robie1995}\\
\ce{MnTiO3(s)} & Pyrophanite & \citet{Robie1995}\\
\ce{Ca2MgSi2O7(s)} & Akermanite & \citet{Robie1995}\\
\ce{MnS(s)} & Alabandite & \citet{Robie1995}\\
\ce{MgCr2O4(s)} & Magnesiochromite & \citet{Robie1995}\\
\ce{CaMgSi2O6(s)} & Diopside & \citet{Robie1979}\\
\ce{Al2O3(s)} & Corundum & NIST-JANAF\\
\ce{Ca3Al2Si3O12(s)} & Grossular & \citet{Robie1995}\\
\ce{C(s)} & Diamond & \citet{Robie1995}\\
\ce{AlN(s)} & Aluminium nitride & NIST-JANAF\\
\ce{Mg3Si2O9H4(s)} & Chrysotile & \citet{Robie1995}\\
\ce{CaS(s)} & Calcium sulfide & NIST-JANAF\\
\ce{Mg2SiO4(s)} & Forsterite & NIST-JANAF\\
\ce{MgO2H2(s)} & Magnesium Hydroxide & NIST-JANAF\\
\ce{MgS(s)} & Magnesium sulfide & NIST-JANAF\\
\ce{MgSiO3(s)} & Enstatite & NIST-JANAF\\
\ce{NaAlSi3O8(s)} & Lingunite & \citet{Robie1995}\\
\ce{CaTiSiO5(s)} & Titanite & \citet{Robie1995}\\
\ce{Ni(s)} & Nickel & NIST-JANAF\\
\ce{FeSiO3(s)} & Ferrosilite & \citet{Robie1995}\\
\ce{FeS(s)} & Troilite & NIST-JANAF\\
\ce{Fe3O4(s)} & Magnetite & NIST-JANAF\\
\ce{Fe3C(s)} & Cementite & \citet{Robie1995}\\
\ce{P(s)} & Phosphorus & NIST-JANAF\\
\ce{Fe2SiO4(s)} & Fayalite & \citet{Robie1995}\\
\ce{Fe(s)} & Iron & NIST-JANAF\\
\ce{Si(s)} & Silicon & NIST-JANAF\\
\ce{Cr2FeO4(s)} & Chromite & \citet{Robie1995}\\
\ce{SiC(s)} & Silicon carbide & NIST-JANAF\\
\ce{Cr(s)} & Chromium & NIST-JANAF\\
\ce{MgAl2O4(s)} & Spinel & NIST-JANAF\\
\ce{Ti2O3(s)} & Titanium(III) oxide & NIST-JANAF\\
\ce{TiC(s)} & Titanium carbide & NIST-JANAF\\
\ce{TiN(s)} & Titanium nitride & NIST-JANAF\\
\ce{CaTiO3(s)} & Perovskite & \citet{Robie1995}\\
\ce{TiO2(s)} & Titanium dioxide & NIST-JANAF\\
\ce{Al2S3(s)} & Aluminium Sulfide & NIST-JANAF\\
\ce{C(s)} & Graphite & \citet{Robie1995}\\
\ce{CaO(s)} & Calcium oxide & NIST-JANAF\\
\ce{Ti4O7(s)} & Tetratitanium Heptoxide & \citet{Robie1979}\\\hline
 \end{tabular}
 \label{tab:moleculeDB_s}
\end{table}

\newpage

\begin{longtable}{lll}
\caption{Included gas-phase molecule data.}\\
\hline\hline
 formula & common name & data source \\\hline
 \endfirsthead
 \caption{continued.}\\
 \hline
formula & common name & data source \\
\hline
\endhead
\hline
\endfoot
\hline
\endlastfoot
\ce{Al(g)} & Aluminium & NIST-JANAF\\
\ce{SiN(g)} & Silicon nitride & NIST-JANAF\\
\ce{SiH(g)} & Silylidyne & NIST-JANAF\\
\ce{SiC(g)} & Silicon carbide & NIST-JANAF\\
\ce{Si(g)} & Silicon & NIST-JANAF\\
\ce{S2(g)} & Sulfur & NIST-JANAF\\
\ce{S(g)} & Sulfur & NIST-JANAF\\
\ce{PS(g)} & Phosphorus sulfide & NIST-JANAF\\
\ce{PO(g)} & Phosphorus oxide & NIST-JANAF\\
\ce{PN(g)} & Phosphorus nitride & NIST-JANAF\\
\ce{PH(g)} & Phosphinidene & NIST-JANAF\\
\ce{P(g)} & Phosphorus & NIST-JANAF\\
\ce{O2(g)} & Oxygen & NIST-JANAF\\
\ce{O(g)} & Oxygen & NIST-JANAF\\
\ce{NS(g)} & Nitrogen sulfide & NIST-JANAF\\
\ce{NO(g)} & Nitrogen oxide & NIST-JANAF\\
\ce{NiS(g)} & Nickel sulfide & NIST-JANAF\\
\ce{Ni(g)} & Nickel & NIST-JANAF\\
\ce{NH3(g)} & Ammonia & NIST-JANAF\\
\ce{NaOH(g)} & Sodium hydroxide & NIST-JANAF\\
\ce{NaO(g)} & Sodium oxide & NIST-JANAF\\
\ce{Na2(g)} & Sodium & NIST-JANAF\\
\ce{Na(g)} & Sodium & NIST-JANAF\\
\ce{N2(g)} & Nitrogen & NIST-JANAF\\
\ce{N(g)} & Nitrogen & NIST-JANAF\\
\ce{MgS(g)} & Magnesium sulfide & NIST-JANAF\\
\ce{MgOH(g)} & Magnesium hydroxide & NIST-JANAF\\
\ce{MgO(g)} & Magnesium oxide & NIST-JANAF\\
\ce{SiO(g)} & Silicon oxide & NIST-JANAF\\
\ce{MgN(g)} & Magnesium nitride & NIST-JANAF\\
\ce{SiS(g)} & Silicon sulfide & NIST-JANAF\\
\ce{SO2(g)} & Sulfur dioxide & NIST-JANAF\\
\ce{Mn(g)} & Manganese & NIST-JANAF\\
\ce{CrO2(g)} & Chromium Oxide & NIST-JANAF\\
\ce{Ca2(g)} & Calcium & NIST-JANAF\\
\ce{COS(g)} & Carbon Oxide Sulfide & NIST-JANAF\\
\ce{PO2(g)} & Phosphorus Oxide & NIST-JANAF\\
\ce{PH3(g)} & Phosphine & NIST-JANAF\\
\ce{PH2(g)} & Phosphino & NIST-JANAF\\
\ce{P4O6(g)} & Phosphorus Oxide & NIST-JANAF\\
\ce{P2(g)} & Phosphorus & NIST-JANAF\\
\ce{HAlO(g)} & Aluminium Hydride Oxide & NIST-JANAF\\
\ce{Al2O2(g)} & Aluminium Oxide & NIST-JANAF\\
\ce{NaCN(g)} & Sodium Cyanide & NIST-JANAF\\
\ce{Na2O2H2(g)} & Sodium Hydroxide & NIST-JANAF\\
\ce{AlO2H(g)} & Aluminium Hydroxide & NIST-JANAF\\
\ce{CaO2H2(g)} & Calcium Hydroxide & NIST-JANAF\\
\ce{MgO2H2(g)} & Magnesium Hydroxide & NIST-JANAF\\
\ce{FeO2H2(g)} & Iron Hydroxide & NIST-JANAF\\
\ce{OH(g)} & Hydroxyl & NIST-JANAF\\
\ce{NH2(g)} & Amidogen & NIST-JANAF\\
\ce{NH(g)} & Imidogen & NIST-JANAF\\
\ce{CH2(g)} & Methylene & NIST-JANAF\\
\ce{PCH(g)} & Methinophosphide & NIST-JANAF\\
\ce{SiO2(g)} & Silicon Oxide & NIST-JANAF\\
\ce{SiH4(g)} & Silane & NIST-JANAF\\
\ce{TiO2(g)} & Titanium dioxide & NIST-JANAF\\
\ce{TiO(g)} & Titanium oxide & NIST-JANAF\\
\ce{Ti(g)} & Titanium & NIST-JANAF\\
\ce{SO(g)} & Sulfur oxide & NIST-JANAF\\
\ce{MgH(g)} & Magnesium hydride & NIST-JANAF\\
\ce{NaH(g)} & Sodium hydride & NIST-JANAF\\
\ce{CP(g)} & Carbon phosphide & NIST-JANAF\\
\ce{Al2O(g)} & Aluminium(I) oxide & NIST-JANAF\\
\ce{AlH(g)} & Aluminium hydride & NIST-JANAF\\
\ce{Fe(g)} & Iron & NIST-JANAF\\
\ce{CS(g)} & Carbon sulfide & NIST-JANAF\\
\ce{CrO(g)} & Chromium oxide & NIST-JANAF\\
\ce{CrN(g)} & Chromium nitride & NIST-JANAF\\
\ce{AlO(g)} & Aluminium(II) oxide & NIST-JANAF\\
\ce{AlOH(g)} & Aluminium hydroxide & NIST-JANAF\\
\ce{FeO(g)} & Iron oxide & NIST-JANAF\\
\ce{AlS(g)} & Aluminium sulfide & NIST-JANAF\\
\ce{Cr(g)} & Chromium & NIST-JANAF\\
\ce{CO2(g)} & Carbon dioxide & NIST-JANAF\\
\ce{CO(g)} & Carbon monoxide & NIST-JANAF\\
\ce{CN(g)} & Cyanogen & NIST-JANAF\\
\ce{Ca(g)} & Calcium & NIST-JANAF\\
\ce{CH4(g)} & Methane & NIST-JANAF\\
\ce{CaS(g)} & Calcium sulfide & NIST-JANAF\\
\ce{CaOH(g)} & Calcium hydroxide & NIST-JANAF\\
\ce{C(g)} & Carbon & NIST-JANAF\\
\ce{FeS(g)} & Iron sulfide & NIST-JANAF\\
\ce{CaO(g)} & Calcium oxide & NIST-JANAF\\
\ce{HS(g)} & Mercapto & NIST-JANAF\\
\ce{Mg(g)} & Magnesium & NIST-JANAF\\
\ce{He(g)} & Helium & NIST-JANAF\\
\ce{H(g)} & Hydrogen & NIST-JANAF\\
\ce{HCO(g)} & Formyl & NIST-JANAF\\
\ce{H2(g)} & Hydrogen & NIST-JANAF\\
\ce{HCN(g)} & Hydrogen cyanide & NIST-JANAF\\
\ce{H2S(g)} & Hydrogen sulfide & NIST-JANAF\\
\ce{H2O(g)} & Water & NIST-JANAF\\\hline
\end{longtable}

\vspace{1cm}

\begin{table}[ht]
\caption{Included reference state molecule data.}
 \centering
 \begin{tabular}{lll}
 \hline\hline
 formula & common name & data source \\\hline
\ce{Al} & Aluminium & NIST-JANAF \\
\ce{Ni} & Nickel & NIST-JANAF \\
\ce{O2} & Oxygen & NIST-JANAF \\
\ce{C} & Carbon & NIST-JANAF \\
\ce{P} & Phosphorus & NIST-JANAF \\
\ce{Ca} & Calcium & NIST-JANAF \\
\ce{N2} & Nitrogen & NIST-JANAF \\
\ce{Ti} & Titanium & NIST-JANAF \\
\ce{Fe} & Iron & NIST-JANAF \\
\ce{Mg} & Magnesium & NIST-JANAF \\
\ce{Cr} & Chromium & NIST-JANAF \\
\ce{H2} & Hydrogen & NIST-JANAF \\
\ce{S} & Sulfur & NIST-JANAF \\
\ce{Si} & Silicon & NIST-JANAF \\
\ce{Na} & Sodium & NIST-JANAF \\\hline
 \end{tabular}
 \label{tab:moleculeDB_ref}
\end{table}

\end{appendix}

\end{document}